\documentclass[fleqn,usenatbib]{mnras}

\usepackage{newtxtext,newtxmath}

\usepackage[T1]{fontenc}

\DeclareRobustCommand{\VAN}[3]{#2}
\let\VANthebibliography\thebibliography
\def\thebibliography{\DeclareRobustCommand{\VAN}[3]{##3}\VANthebibliography}

\usepackage{graphicx}	
\usepackage{amsmath}	

\newcommand{\Msun}{M$_{\odot}$}
\newcommand{\Msunmath}{\text{M}_{\odot}}
\newcommand{\HST}{\textit{HST}}

\newcommand\numberthis{\addtocounter{equation}{1}\tag{\theequation}}

\newcommand{\review}[1]{#1}

\title[SMBH lensing in cluster field]{Gravitational lensing effects of supermassive black holes in cluster environments}

\author[G. Mahler et al.]{
Guillaume Mahler,$^{1,2}$\thanks{E-mail: guillaume.mahler@durham.ac.uk}
Priyamvada Natarajan,$^{3,4,5}$
Mathilde Jauzac,$^{1,2,6,7}$
Johan Richard,$^{8}$  
\\
$^{1}$Institute for Computational Cosmology, Durham University, South Road, Durham DH1 3LE, UK\\
$^{2}$Centre for Extragalactic Astronomy, Durham University, South Road, Durham DH1 3LE, UK\\
$^{3}$Department of Astronomy, Yale University, 52 Hillhouse Avenue, New Haven, CT 06511, USA\\
$^{4}$Department of Physics, Yale University, P.O. Box 208121, New Haven, CT 06520, USA\\
$^{5}$Black Hole Initiative, Harvard University, 20 Garden Street, Cambridge, MA 02138, USA\\
$^{6}$Astrophysics Research Centre, University of KwaZulu-Natal, Westville Campus, Durban 4041, South Africa\\
$^{7}$School of Mathematics, Statistics \& Computer Science, University of KwaZulu-Natal, Westville Campus, Durban 4041, South Africa\\
$^{8}$Univ Lyon, Univ Lyon1, Ens de Lyon, CNRS, Centre de Recherche Astrophysique de Lyon UMR5574, 69230, Saint-Genis-Laval,
France
}

\date{Accepted XXX. Received YYY; in original form ZZZ}

\pubyear{2022}

\begin{document}
\label{firstpage}
\pagerange{\pageref{firstpage}--\pageref{lastpage}}
\maketitle

\begin{abstract}
This study explores the gravitational lensing effects of supermassive black holes (SMBHs) in galaxy clusters. While the presence of central SMBHs in galaxies is firmly established, recent work from high-resolution simulations predict the existence of an additional population of wandering SMBHs. Though the masses of these SMBHs are a minor perturbation on the larger scale and individual galaxy scale dark matter components in the cluster, they can impact statistical lensing properties and individual lensed image configurations. Probing for these potentially observable signatures, we find that SMBHs imprint detectable signatures in rare, higher-order strong lensing image configurations although they do not manifest any statistically significant detectable evidence in either the magnification distribution or the integrated shear profile. Investigating specific lensed image geometries, we report that a massive, near point-like, potential of an SMBH causes the following detectable effects: (i) image splitting leading to the generation of extra images; (ii) positional and magnification asymmetries in multiply imaged systems; and (iii) the apparent disappearance of a lensed counter-image. Of these, image splitting inside the cluster tangential critical curve, is the most prevalent notable observational signature. We demonstrate these possibilities in two cases of observed giant arcs in SGAS\,J003341.5+024217 and RX\,J1347.5-1145, wherein specific image configurations seen can be reproduced with SMBHs. Future observations with high-resolution instrumentation (e.g. MAVIS-Very Large Telescope, MICADO-Extremely Large Telescope, and the upgraded ngVLA, along with data from the \textit{Euclid} \& \textit{Nancy Grace Roman} Space Telescopes and the Rubin LSST Observatory are likely to allow us to probe these unique yet rare SMBHs lensing signatures.
\end{abstract}





\section{Introduction}

Gravitational lensing has emerged as a powerful method to probe the detailed mass distribution on multiple cosmic scales in the Universe, from individual galaxies, groups of galaxies to clusters of galaxies, which all serve as effective lenses for the distant background galaxies and quasars (see review by \citealt{KneibNatarajan2011} for details). We now know that most, if not all, galaxies in the Universe likely harbor a central supermassive black hole (SMBH). SMBHs are ubiquitous at the centers of galaxies, and the most massive ones are expected in the brightest cluster galaxies (BCGs) that anchor the center of the gravitational potential well in galaxy clusters. Observations suggest that properties of central SMBHs are correlated to properties of their host galaxy \citep{Magorrian+1998,Gebhardt2000,Tremaine+2002,Ferrarese&Merritt2000}. Even though the mass of the central SMBH is negligible compared to the mass of the stellar component of its host galaxy, in just the bulge $M_{\rm bh} \sim 10^{-3} M_{\rm bulge}$, coupling scales the SMBH nevertheless appears to play an important role in modulating star formation in the galactic nucleus.\\

Gravitational lensing, predicted by General Relativity results in the deflection of light paths by strong gravitational potentials encountered en-route. Light from distant background galaxies is deflected by the foreground potential of a galaxy cluster and its member galaxies producing a multiplicity of detectable effects. Gravitational lensing observations provide strong constraints on the inner density profiles of galaxies and clusters due to the production of multiple images and highly magnified distorted arcs. In this paper, we explore the gravitational lensing effects produced by SMBHs in cluster environments. 

Lensing theory predicts that \review{for simple isolated galaxy lenses} every strong lensing system should produce a faint, demagnified image at the \review{very inner} center of the lens
However, these central images \review{have only been hinted \citep{Winn2004} in lensed radio galaxies and multiple} studies have focused on them as there is scarce contamination by the lens itself (e.g. \citealt{Winn2004,Rusin2005,Wong2015,Tamura2015,Quinn2016,Wong2017}). \\

Realistic models of galaxies with a dominant central stellar component predict a wide range of properties for these core images spanning a range of magnification factors. Despite systematic searches for these central de-magnified images, they have not been found. While tweaking properties of the stellar component can account for the absence of these images, \cite{Mao2001} showed that the presence of a central SMBH introduces new qualitative features in the resulting critical and caustic curves that could easily destroy the presence of these central images. \\

Traditionally, the detection of active SMBHs has been through X-ray studies of the accretion process that has permitted mass measurements. For nearby, dormant SMBHs, their masses have been successfully constrained mapping the gravitational potential using stars and gas when available as tracers. However, the kind of data needed for dynamical modeling to determine SMBH masses is unavailable for sources beyond 50-100\,Mpc. Since lensing is achromatic, mapping the shadow of gas accretion including lensing effects close to the horizon, recently permitted the Event Horizon Telescope (EHT) Collaboration to garner a mass measurement for the SMBH at the center of M\,87. The central SMBH in M\,87 is reported to have a mass of $6.5 \times 10^9\Msunmath$ \cite{EHT2019}. While such measurements are infeasible for $z > 0.1$ SMBHs, lensing effects resolvable with the next generation of interferometric radio arrays like the ngVLA might provide an entirely new method to measure SMBH masses. It is this, in part, that motivates our current study. In particular, if any unique detectable SMBH lensing signatures exist, they would offer a novel way to find dormant SMBHs at inter-mediate redshifts, $0.2 < z < 1.0$, and serve as invaluable addition to our understanding of the growth and evolution black hole populations over cosmic time.\\

Previous theoretical work has focused on including the presence of a central SMBH in the galactic nucleus of an individual galaxy lens. The gravitational potential of an SMBH was added to various assumed galaxy mass profiles, ranging from a cored isothermal sphere \citep{Mao2001} and a Plummer model \citep{Werner2006} to explore their combined lensing effects. Recently, \citet{Karamazov2021a} include the central SMBH as a point mass embedded in an NFW profile to model the BCG in a fiducial galaxy cluster. \\

Given the clear cut prediction of the absence of the central image \review{in the case of isolated galaxies}, various groups including \citet{Keeton2003}, \citet{Chen2018}, and \citet{Inada2008} have looked at optical and radio data of distant galaxies (z$\sim$0.2-1) to search for the central image. Multiple studies have been focused on lensed radio galaxies as there is no contamination  by the lens itself (e.g. \citealt{Winn2004,Rusin2005,Wong2015,Tamura2015,Quinn2016,Wong2017}). While these de-magnified images remain undetected, this absence of detection in itself is not a clear cut signal telegraphing the presence of a central SMBH. In fact, \citep{Smith2017} showed that a change in the stellar IMF can also alter the central density of the stellar component in the inner regions of galaxies sufficiently to remove central images. Therefore, the absence of a central image cannot be assumed to provide smoking gun evidence for an SMBH. \review{Central images could appear more clearly in lensing potential embedded in more complex environment as it could shift demagnification region outside the inner core of the galaxy (e.g. \citealt{Dahle2013, Sharon2017,Ostrovski2018,Muller2020,Martinez2022})}\\ 

Additionally, the presence of dark substructures (whose lensing effects would be degenerate with that of SMBHs) have also been inferred from galaxy-galaxy lensing studies, as reported in \citet{Minor2021a} and \citet{Minor2021b} where they argue for evidence of the existence of massive compact sub-halos perturbing the observed lensing signal. In the latter case, an associated, luminous counterpart is seen but that could also be related to additional sources lensed behind \citep{Collett2020,Smith2021}.

In the work presented in this paper, we extend and expand previous studies to examine in detail the lensing signatures SMBHs in a range of cluster environments: the case of a central SMBH in a cluster BCG; a central SMBH in a cluster member galaxy, and the more general case of a wandering SMBH in the cluster environment. We are motivated to study the effect of this additional wandering SMBH population due to recent work by \cite{Ricarte2021a,Ricarte+2021b}. Analyzing a high-resolution simulated cluster, Romulus-C, \citep{Ricarte2021a} reported the existence of a large population of wandering SMBHs in cluster environments originating from the tidal stripping and disruption of in-falling dwarf galaxies. In addition, microlensing in clusters is very sensitive to the mass but remains a very rare event, though recently \citet{Liang2020} argue that significant monitoring of subtle changes in the magnification would be able to yield accurate mass measurements leading to the potential implication of IMBHs or SMBHs. Previous reports using microlensing events by stars in clusters (e.g. Icarus and Iapix in MACS\,J1149 -- \citealt{Kelly2018,Diego2018}) have been searched for. \\

The plan of our paper is as follows: in Section~\ref{sec:model}, we present the methods used to model and study the lensing effects of SMBHs in clusters; the lensing effects of the central BCG SMBHs, the central SMBHs in cluster galaxies and wandering SMBHs are explored in Sections~\ref{sec:BCGcentral}, \ref{sec:central} and \ref{sec:wanderers} respectively. The impact of SMBHs on specific lensing image configurations is studied in detail and presented in Section~\ref{sec:obs-effect}, and the observational case studies in current data wherein SMBHs may be implicated are discussed in Section~\ref{sec:realcase}. We conclude in Section~\ref{sec:ccl} with the prospects for future detection and delineation of SMBH lensing with the upcoming data deluge from new facilities.\\

\section{Modeling the lensing effects of SMBHs}\label{sec:model}

Here we look more exhaustively at the range of lensing phenomena that could be revealed with the explicit inclusion of a central SMBH in a cluster galaxy or wandering SMBHs in clusters. We model the gravitational potential of the entire system holistically following \citep{PNKneib1997} as a superposition of the following components: larger scale smooth components (that model the distribution of the smoothly distributed dark matter); a sum over galaxy-scale perturbers (that model the contribution of the dark matter subhalos associated with individual cluster galaxies) and now explicitly include the associated SMBH population (those hosted at the centers of cluster galaxies or as wanderers) as:

\begin{eqnarray}
\phi\,=\,\phi_{\rm smooth}\,+\,\Sigma_{i}\phi_{\rm clusgal}\,+\,\Sigma_{j}\phi_{\rm BH}
\end{eqnarray}

As detailed below, we use mass profiles whose lensing properties are well understood to model mass components above.

\subsection{Modeling the composite profile}

To allow flexibility in our computation, we  derive lensing properties adopting a dual Pseudo Isothermal Elliptical profile (dPIE; \citealt{Kassiola1993,PNKneib1997, Eliasdottir2007}) often used in lensing analysis. \review{We use this profile to model cluster halos, galaxy halos and SMBHs.} This profile has the advantage that the first and second partial derivatives of the lensing potential can both be written out analytically. In addition, this profile offers a critical free parameter, the core radius, that can be tuned to flatten the central density distribution flexibly, that is of great utility while modeling the effect of SMBHs. 

The 3D density distribution of the dPIE is given by:
\begin{equation}
    \rho(r)=\frac{\rho_0}{(1+\big(\frac{r}{r_{\rm core}}\big)^2)(1+\big(\frac{r}{r_{\rm cut}}\big)^2)} ; r_{cut}>r_{\rm core}.
\end{equation}
Following the details presented in Appendix\,A of \cite{Eliasdottir2007}, we adopt a fiducial central velocity dispersion $\sigma_{\rm{dPIE}}$ to relate to the central density as follows:

\begin{equation}
    \sigma_{\rm{dPIE}}^2=\frac{4}{3}G\pi\rho_{0} \frac{r_{\rm core}^2r_{\rm cut}^3}{(r_{\rm cut}-r_{\rm core})(r_{\rm cut}+r_{\rm core})^2}. \label{eq_sigma}
\end{equation}
The convergence, $\kappa$, and the shear, $\gamma$, of a single dPIE are given by:

\begin{equation}
    \kappa(R)\equiv \frac{\Sigma(R)}{\Sigma_{\rm crit}}=\frac{\Sigma_0}{\Sigma_{\rm crit}}\frac{r_{\rm core}r_{\rm cut}}{r_{\rm cut}-r_{\rm core}} \left( \frac{1}{\sqrt{r_{\rm core}^2+R^2}}-\frac{1}{\sqrt{r_{\rm cut}^2+R^2}}\right),
\end{equation}
and

\begin{align*}
 \gamma(R)= & \frac{\Sigma_0}{\Sigma_{\rm crit}}\frac{r_{\rm core}r_{\rm cut}}{r_{\rm cut}-r_{\rm core}} \\
 & \left[2\left(\frac{1}{r_{\rm core}+\sqrt{r_{\rm core}^2+R^2}}-\frac{1}{r_{\rm cut}+\sqrt{r_{\rm cut}^2+R^2}}\right) + \right. \\
 &\left. \left(\frac{1}{\sqrt{r_{\rm core}^2+R^2}}-\frac{1}{\sqrt{r_{\rm cut}^2+R^2}}\right)\right]
\numberthis \label{eqn}
\end{align*}

with 
\begin{equation}
    \Sigma_{0}=\pi \rho_0 \frac{r_{\rm core}r_{\rm cut}}{r_{\rm cut}+r_{\rm core}}
\end{equation}
\begin{equation}
    \Sigma_{crit}\equiv \frac{c^2}{4\pi G}\frac{D_{\rm S}}{D_{\rm L}D_{\rm LS}}
\end{equation}
where $D_{\rm L}$, $D_{\rm S}$, and $D_{\rm LS}$ are the angular diameter distances from the observer to the lens, the observer to the source, and between the lens and source respectively.
 
\subsection{Explicit inclusion of the SMBH component}

We define the black hole mass hosted in a cluster galaxy by adopting the fiducial velocity dispersion of a dPIE profile as the central velocity of the bulge of the host galaxy. We can therefore, use the well known empirically derived local black hole mass - bulge mass relation \citep{Gultekin2009}: 

\begin{equation}
 Log(\frac{M_{\rm bh}}{M_{\odot}}) = \alpha + \beta * Log(\frac{\sigma_{\rm bulge}}{200\,{\rm  km\,s^{-1}}}),
\label{eq:M-sigma}
\end{equation}
where $\alpha$ = 8.12, $\beta$ = 4.24 and $\sigma_{\rm bulge}$ is the bulge velocity dispersion defined here as the dPIE velocity dispersion shown in equation \ref{eq_sigma}.

\begin{figure}
    \centering
    \includegraphics[width=\linewidth]{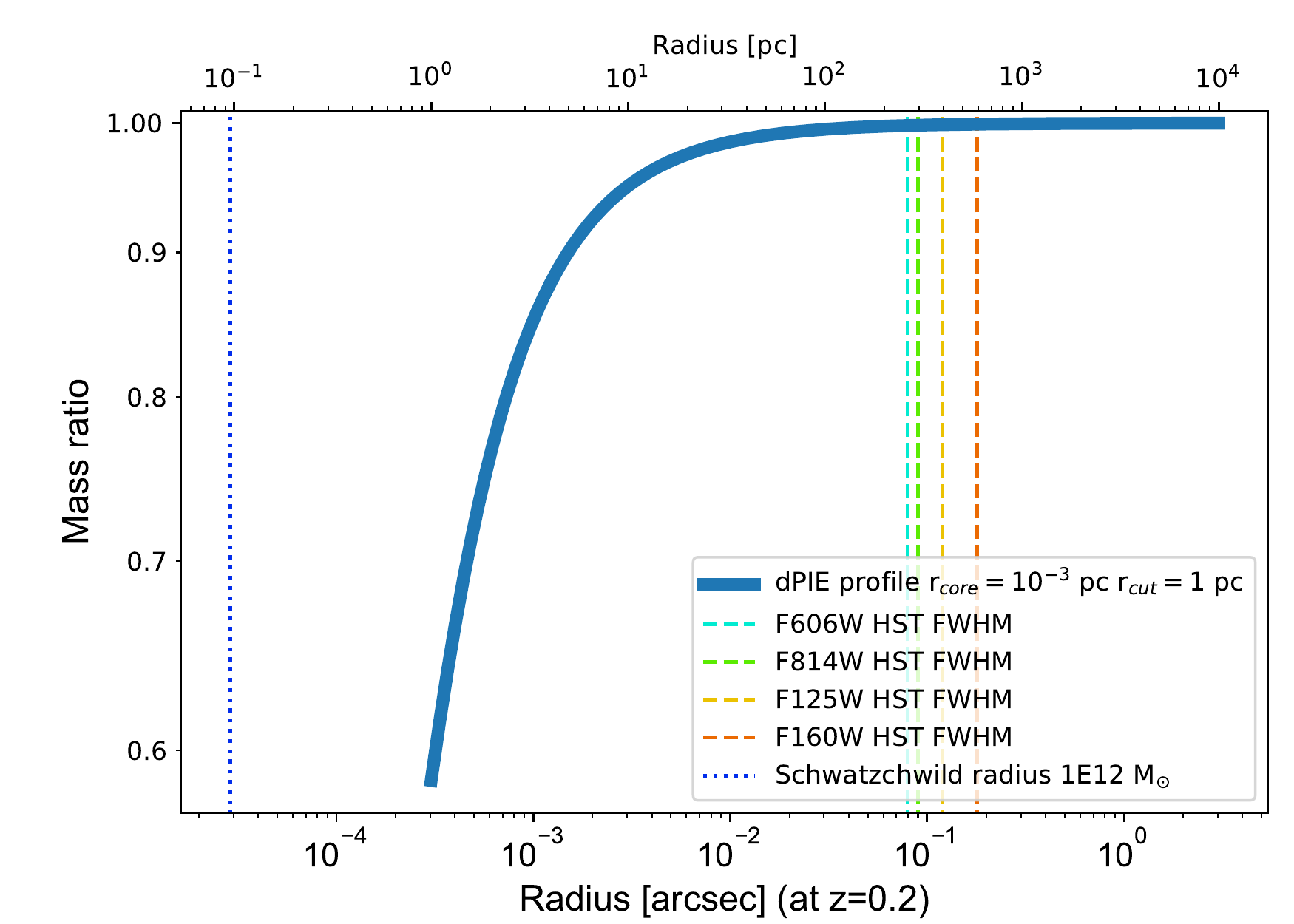}
    \includegraphics[width=\linewidth]{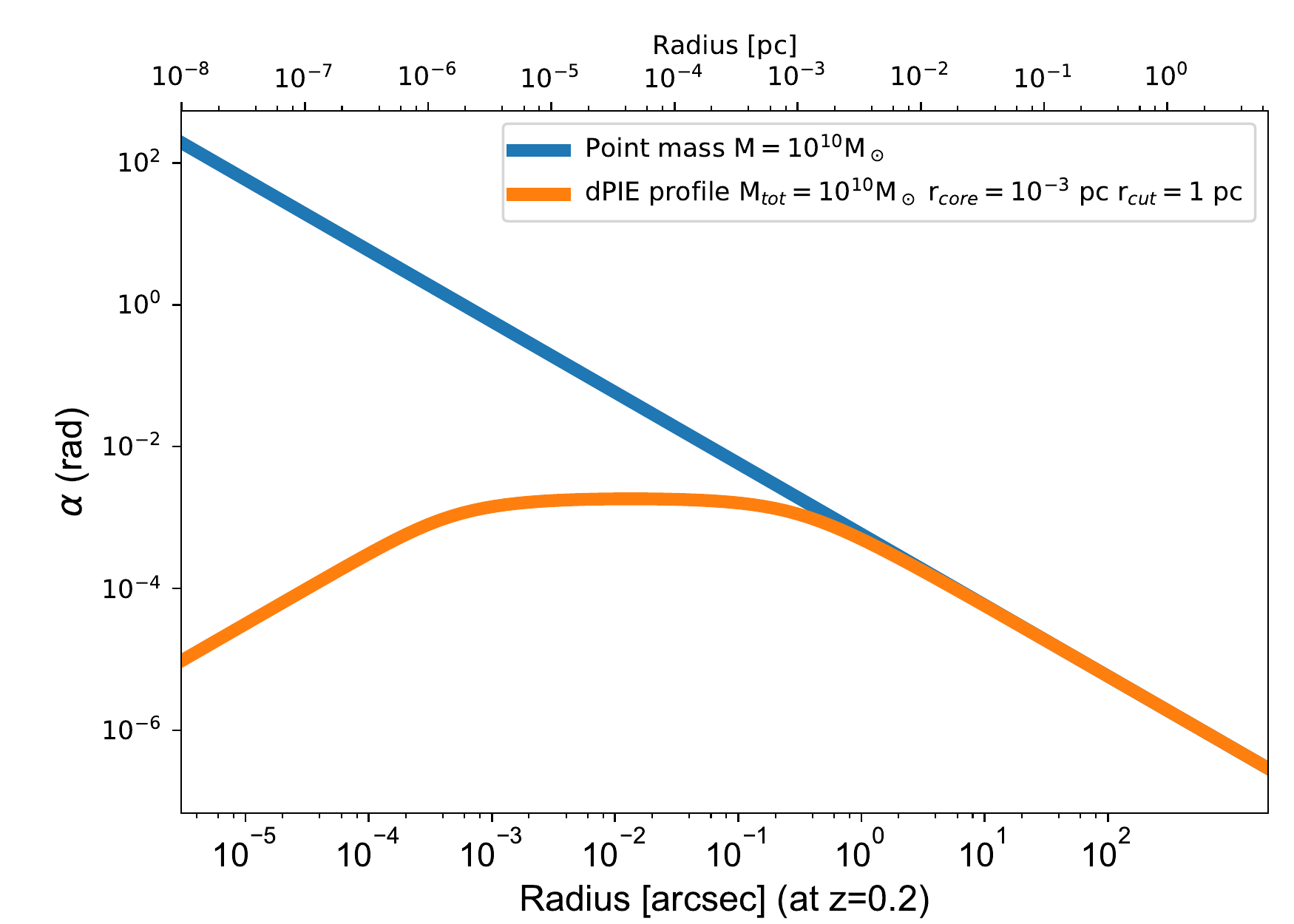}
    \caption{
    Top: Mass profile of a dPIE potential mimicking a SMBH normalised by the total mass with a cut radius of $r_{\rm cut}=1$\,pc and a core radius of $r_{\rm core}=1\times10^{-3}$\,pc. 
    The dotted line corresponds to the Schwarzschild radius of a $10^{12}\Msunmath$ SMBH, and the dashed line corresponds to \emph{HST} FWHM of typical filter from measurements made by the CANDELS collaboration \url{https://irsa.ipac.caltech.edu/data/COSMOS/images/candels/hlsp_candels_hst_cos-tot_readme_v1.0.pdf}. 
    \review{Bottom: The deflection angle $\alpha$ in function of radius for a point mass and our dPIE formalism. We can see a complete overlap after 1 arcsecond  lens a  profile}
    } 
    \label{fig:M-r}
\end{figure}

We then correlate the mass of a central SMBH to the dPIE profile to compute the lensing effect of the SMBH. Using a circularly symmetric profile, we adopt a cut radius of $r_{\rm cut}=1.0$\,pc, and a core radius of $r_{\rm core}=0.001$\,pc.
\review{These values are taken to account for an small accretion disks around the SMBH but its size is negligible for the current observations considered later in this analysis}
\review{The top panel of} Figure\,\ref{fig:M-r} shows the integrated mass profile of such a potential normalized by its total mass. This corresponds to roughly 99.8\% of the mass within the 330\,pc corresponding to $\sim$0.1\arcsec\ at $z = 0.2$, which adequately captures the gravitational potential of an SMBH. \review{The bottom panel shows the deflection angle computed for our profile and a point mass, we can see that after 0.01\,pc ($\sim$1\arcsec\ at $z = 0.2$) the deflection angle is identical to a point mass.}

We compute the shear and convergence profile from this compact dPIE to demonstrate that it is an appropriate representation of the point mass.  We find that the kappa profile, following the density profile presented in Figure 1, reaches $\sim$0.1 after 0.02\arcsec - extremely similar to the lensing behavior produced by the analytic point mass profile used by \citet{Karamazov2021a,Karamazov2021b}. We remind the reader that the convergence for a point mass is zero everywhere except at its location. The shear profile for a point mass and our dPIE are nearly identical to < 1\% after 20\,pc (or 0.001\arcsec for a lens at $z = 0.2$). Therefore, we argue that our choice of dPIE profile offers a reasonable and robust representation to model both a central and a wandering SMBH with an associated residual stellar component resulting from the tidal stripping/merging of an infalling galaxy into the cluster environment.

Within this setting, the mass of the dPIE profile is set entirely by the velocity dispersion, $\sigma$. Following the formula derived from \cite{Eliasdottir2007} equations A11 and A25, we have:
\begin{equation}
    \sigma_{dPIE}^{\rm SMBH} = \sqrt{ M_{\rm bh} * \frac{4G}{6\pi} * \frac{r_{\rm cut}}{(r_{\rm cut}^2-r_{\rm core}^2)} } \approx \sqrt{ \frac{M_{\rm bh}}{r_{\rm cut}} * \frac{4G}{6\pi} }
\end{equation}

\section{Lensing by the central BCG SMBH}
\label{sec:BCGcentral}

We first examine the lensing effects produced by the massive SMBH hosted at the center of the cluster BCG. As noted by \cite{Karamazov2021b}, the inclusion of a point like mass at the center of the cluster BCG potential can strongly affect the resulting lensing configurations. \cite{Karamazov2021a} distinguish two regimes where formally the ratio of the convergence produced by the BCG to that of the SMBH exceeds $10^{-4}$, when a single NFW profile is used to model the BCG and a point-like source the SMBH. 

\begin{figure}
    \centering
    \includegraphics[width=\linewidth,trim={22cm 0 22cm 0 },clip]{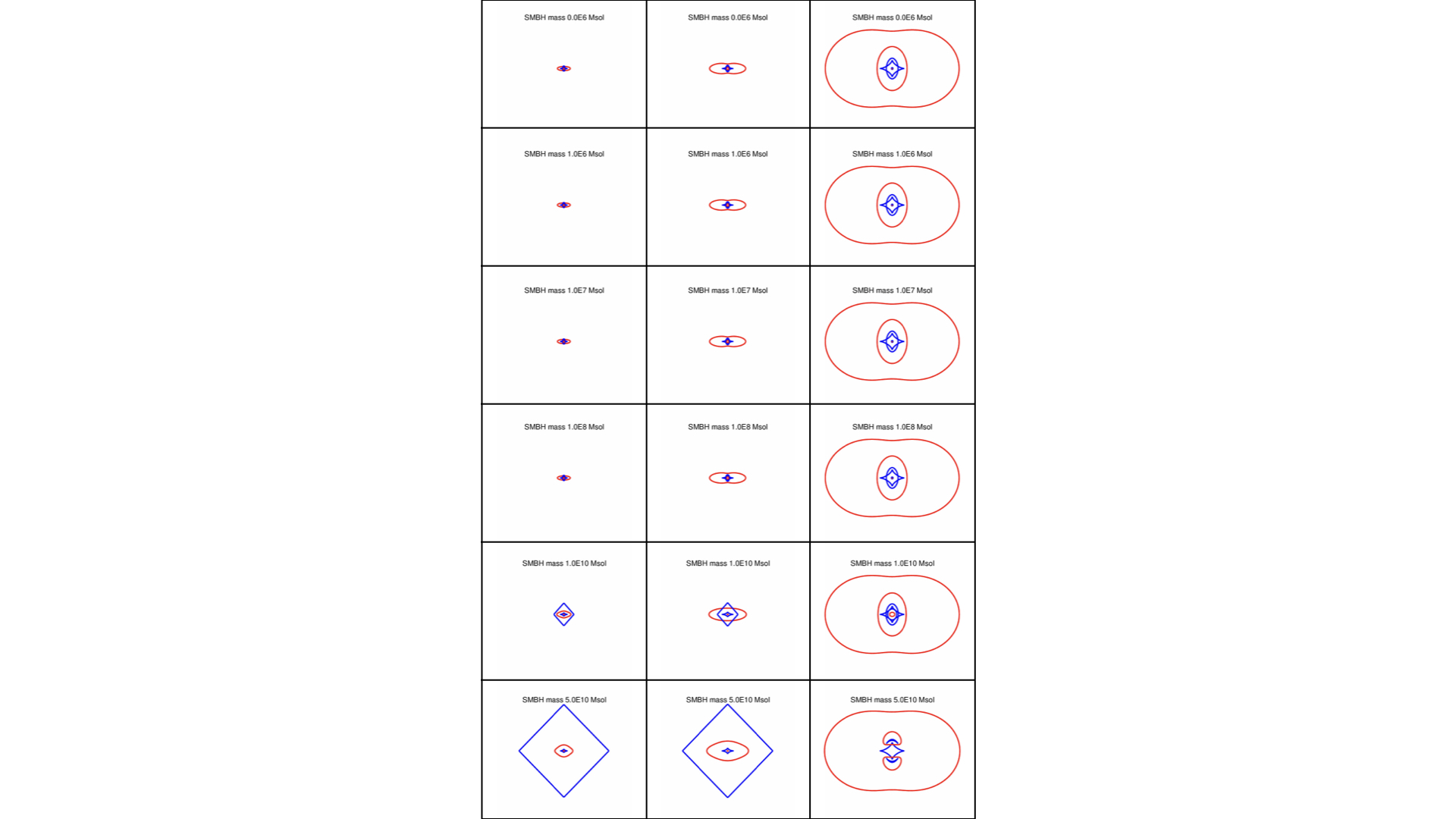}
    \caption{Lensing configurations for a galaxy \review{with a total mass of $10^{12}\Msunmath$ (r$_{{\rm core}}=0.1$\,kpc, r$_{{\rm cut}}=50$\,kpc, $\sigma_{\rm{dPIE}}=135$km\,s$^{-1}$ ellipticity=0.8)} hosting a central SMBH with varying mass (mass increasing from top to bottom) in different environments. \textit{Left}: isolated field galaxy - \textit{middle}: dPIE in a cluster environment with the larger scale component characterized by \review{ellipticity=0.1}, r$_{{\rm core}}=30$\,kpc, r$_{{\rm cut}}=1500$\,kpc, $\sigma_{\rm{dPIE}}=552$km\,s$^{-1}$ yielding a total mass  M$_{tot}=5.10^{14}\Msunmath$  \textit{right}: SMBH embedded in a larger scale cluster modeled with an NFW profile characterized by the following parameters: $c=7$, r$_s=200$\,kpc,  \review{ellipticity=0.1} and a total mass of M$_{200}=3.8~10^{14}\Msunmath$. It is seen that at specific mass the SMBH embedded in an NFW profile splits the internal critical curve (two lowest panels of the right column). This transition occurs in this configuration for a mass of M$_{\rm SMBH}= 3.50~10^{10}$\Msun. Each box has a length of 20\arcsec aside, corresponding to 66\,kpc at z $=0.2$. Further explorations of the lensing configurations produced by a range of SMBH masses are available at: \url{https://sites.google.com/view/guillaume-mahler-astronomer/paper-animation/paper-animation-central-smbh} }
    \label{fig:BCG-panel}
\end{figure}

Our empirical, observationally-based approach does not make such distinctions of regimes based on the convergence, and instead we focus on the ratio of measured quantities - the mass ratio between different contributors to the overall mass budget. Highlighted in Figure\,\ref{fig:BCG-panel} we show how significantly a point-like mass can affects the lensing configuration that is produced. For a   host galaxy \review{with a total total mass of $10^{12}\Msunmath$ (as computed following \citealt{Eliasdottir2007})} with an increasing value of central black-hole mass, we show the effect in the left column, for the case of an isolated galaxy (not embedded in the cluster environment); in the middle panel for a galaxy in a smoothed dPIE potential mimicking a central SMBH in a cluster BCG with a total mass of $5~10^{14}\Msunmath$ ($\sigma_{dPIE} = 552 {\rm km s}^{-1},\,{\rm r_{core}}\,=\,30\,kpc, {\rm r_{cut}}\,=\,1500\,kpc$) and in the rightmost panel, the same galaxy now embedded instead in an NFW halo with a total mass of about $3.8~10^{14}\Msunmath$. Increasing the mass of the central SMBH in the galaxy produces the remarkable effect of boosting the radial caustic curve (blue diamond) outside the tangential caustic curve in the case of a dPIE profile (two left columns). In the case of the SMBH at the center of a cluster scale potential (middle column) the area occupied by the radial caustic goes from 0.4 arcsec$^2$ for an SMBH mass of $10^{9}\Msunmath$ to 5 arcsec$^2$ for an ultra-massive $10^{10}\Msunmath$ SMBH. The outer critical curve is only slightly boosted. A similar boosting effect is not observed for the case of the NFW profile cluster halo (right column). In this case the most noticeable effect that appears is the splitting of the radial critical curves for a massive enough cluster. For the case shown in Figure \ref{fig:BCG-panel} such rare and exotic catastrophic configurations appear only when the SMBH mass is tuned up to reach $M_{\rm SMBH}= 3.50~10^{10}$\Msun (see \citealt{Schneider1992,Orban2009} for more details on rare lensing geometries). \review{The ellipticity of the cluster scale halo can influence the mass of when the splitting of the radial critical curve occurs, as higher elliptical profile increase the density in one direction, we documented two cases for which we vary the ellipticity in Appendix \ref{sec:app-ell}}

\section{Lensing signals of central SMBHs in cluster galaxies}
\label{sec:central}

In this section we study the lensing effect of a SMBH hosted at the center of a cluster member galaxy. Due to the lensing boost from the underlying larger scale cluster components, stronger effects can be produced as explained below. For exploring these effects, we start with modeling the underlying environment with a mock cluster. We add a single cluster scale halo and sub-halos to mimic the contribution of cluster member galaxies. Due to the sensitivity of the lensing signal to the parameters of the mass models, we used a realistic lens model of an observed cluster lens and re-arrange the cluster member locations keeping their radial distances unchanged in multiple realizations. The main dark matter central velocity dispersion and cut radius have been slightly reduced to match a total mass of $5\,10^{14}\,\Msunmath$. The final model used here is available online\footnote{\url{https://sites.google.com/view/guillaume-mahler-astronomer/lens-model}}. We now include a SMBH with mass as expected from the empirical scaling relations at the center of cluster member galaxies. 

\subsection{Effects of central SMBHs on the lensing configurations}

\begin{figure}
    \centering
    \includegraphics[width=\linewidth,trim={23cm 0 23cm 0 },clip]{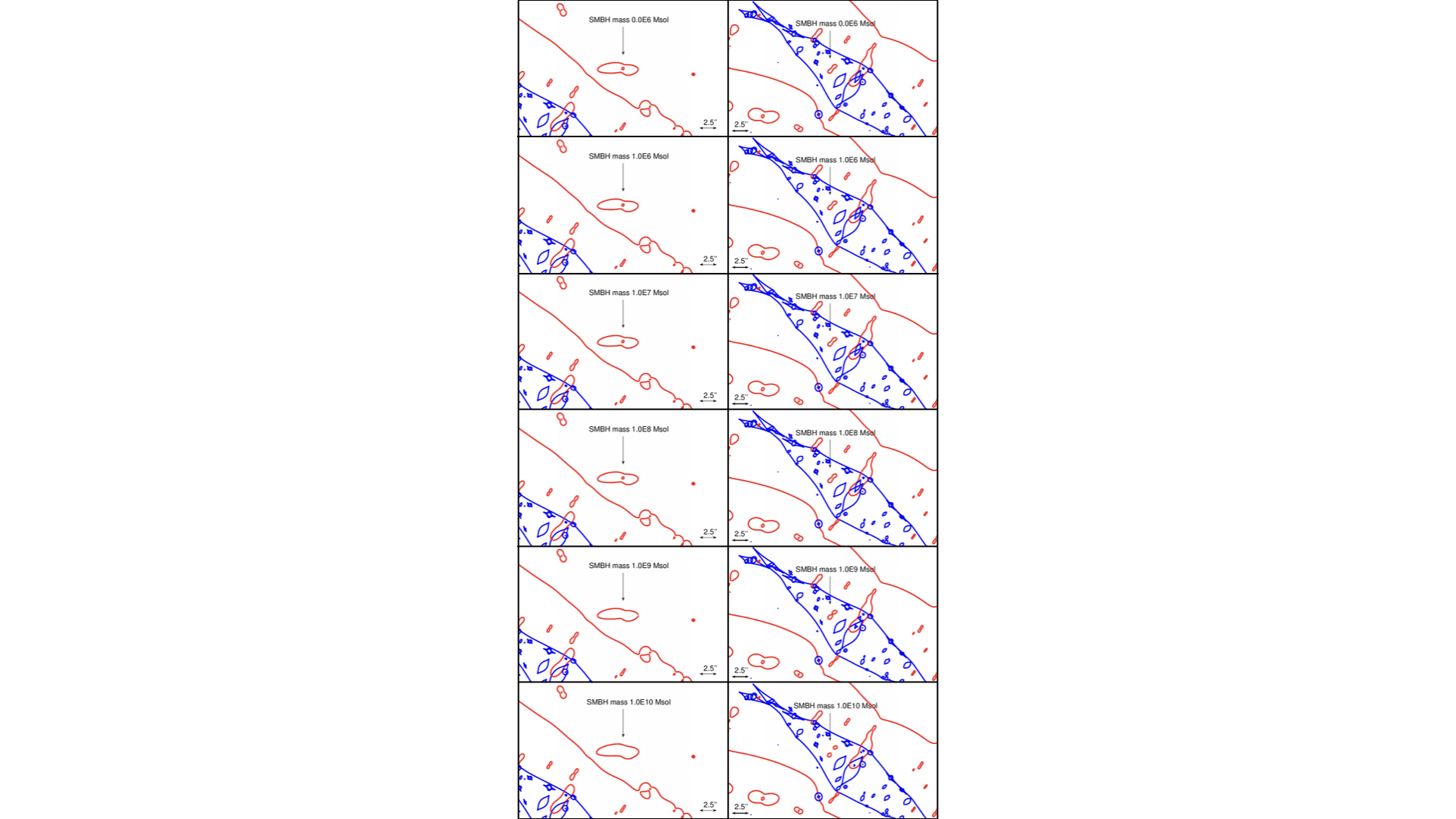}
    \caption{Lensing configurations produced by galaxies with increasing mass of the central SMBH from top to bottom:  0, 10$^{6}$, 10$^{7}$, 10$^{8}$, 10$^{9}$, 10$^{10}$ \Msun. \review{The SMBH is located a the center of each frame, at center of the galaxy pointed by the arrow}
    \textit{Left:} 10$^{12}$ \Msun\, galaxy with central SMBH placed outside the cluster's main critical line. The main effect of a massive SMBH is to remove the central critical curve of the galaxy. 
    \textit{Right:} 10$^{12}$ \Msun\, galaxy placed inside the main critical line of the large scale cluster. In this case the effect of increasing the mass of the SMBH leads to the shrinking of the central part of the galaxy's critical line leading to eventually causing it to split into two. Each box shown is 20\arcsec on a side. Further explorations of the lensing configurations as a function of SMBH masses are available at: \url{https://sites.google.com/view/guillaume-mahler-astronomer/paper-animation/paper-animation-centralsmbhcluster}}
    \label{fig:CM-panel}
\end{figure}

The combined lensing power of the cluster, the host galaxy and its central SMBH are now studied in detail. We identify two main cases and the resulting lensed image geometries. The first case is when the host galaxy lies outside the main cluster critical curve, as illustrated in Figure\,\ref{fig:CM-panel}. In this scenario, the cluster member galaxy produces its own critical curve. The principal effect for such a configuration appears to be the disappearance of the radial critical curve of the galaxy, similar to the effect studied in the case of isolated individual galaxy lenses (e.g. \citealt{Mao2001}). However, due to the lensing boost from the cluster, even less massive galaxies can produce large enough critical curves to potentially produce  detectable image configurations. Secondly, a more spectacular effect appears when the host galaxy is situated inside the main cluster critical curve. In this case, the cluster member galaxy's critical curve serves in practice as the radial critical curve for the overall cluster. Increasing the mass of the SMBH for such configurations, we find, first creates a cassinoid shape (peanut-like) before splitting the curve into two separate rounded critical curves as seen in the left column of Figure\,\ref{fig:CM-panel}. This unique configuration would offer a compelling case to look for a central SMBH in a cluster member galaxy in the inner core region of a cluster-lens. However, upon varying the SMBH mass, we note that the mass needed for such an event to occur is high and in the case displayed Figure\,\ref{fig:CM-panel} this appears for a SMBH at about $10^9\Msunmath$. Such a high mass corresponds to the existence of an over-massive SMBH in a $10^{12}\Msunmath$ host galaxy (an outlier  away from the typical relation \citep{Gultekin2009}, however such cases have been reported \citep{vandenbosch2012}.

\subsection{Statistical effects of central SMBHs on the shear profile of sub-halos}

\begin{figure}
    \centering
    \includegraphics[width=\linewidth]{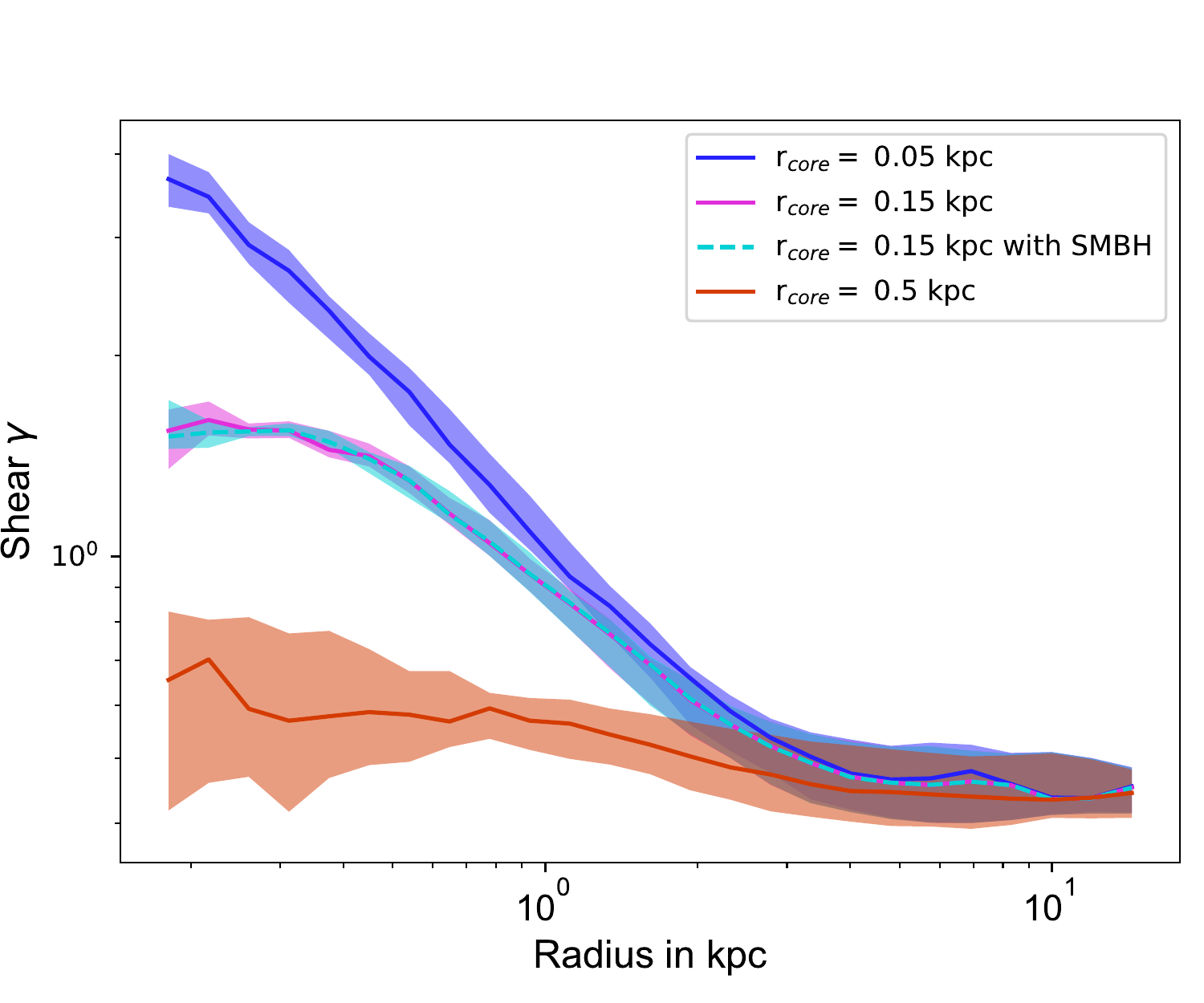}
    \caption{The stacked shear profile of the bright cluster member galaxies: bright cluster members galaxies are defined as those with a magnitude m, in the range $M^*-1<$m$<M^*+1$. The solid lines represent galaxies without a SMBH, but with varying core radius. From top to bottom the core radii are 0.05 kpc, 0.15 kpc, and 0.5 kpc. The dotted line corresponds to the shear profile with a central SMBH hosted by cluster member galaxies. The shaded region represents the 1-sigma asymmetric distribution of shear strength for bright galaxies, with and without a central SMBH.}
    \label{fig:stackgamma}
\end{figure}

Masses of sub-halos in clusters have been detected from statistically combining detected strong and weak lensing signals \citep{Natarajan2017,PastorMira2011,Sifon2018,Niemiec2018}. Future surveys from the LSST Vera Rubin observatory, or the \textit{Nancy Grace Roman Telescope} are expected to overcome statistical limitations by detecting a large population of hitherto undetected massive cluster lenses. We investigate the potential statistical signature of the central SMBHs in stacked shear profiles of cluster member galaxies. We use our mock cluster presented in section \ref{sec:central} associating every cluster member galaxy with a central SMBH following equation \ref{eq:M-sigma}. We produce a shear profile at the location of every cluster member and stack the signal for bright galaxies, corresponding to the luminosity range M$^*-1<$m$_{\rm gal}<$M$^*+1$. We then compare the profiles for the same selected galaxies for the realization without a central SMBH. In addition, we compare the stacks with the signal derived for the known degenerate case altering the internal structure for the cluster galaxies with a modified core core radius, from r$_{\rm core}^*=0.15\,$kpc with the signal using a smaller core r$_{\rm core}^*=0.05\,$kpc and a larger core with r$_{\rm core}^*=0.5$kpc. In our simulation, we use the same scaling relations with luminosity for the core radius r$_{\rm core}=$\,r$_{\rm core}^* (L/L^*)^{1/2}$, as adopted in several previous works by \citep{Jullo2007,Richard2010,Mahler2018,Mahler2019}.  
Figure \ref{fig:stackgamma} shows the stacked shear profiles, where it is clearly seen that the effect on the shear profile arising from changing the core radius of the galaxy far exceeds the effect of adding in a SMBH. Therefore, we conclude that the statistical lensing signature of a central SMBH in the shear profile is degenerate with that of the properties of the stellar component in the inner regions of cluster member galaxies and hence not discernible.

\section{Lensing signals of wandering SMBHs}
\label{sec:wanderers}

In addition to the wandering population of SMBHs detected in the high-resolution simulated cluster Romulus-C that includes the dark matter and the baryon component \citep{Tremmel2019}, dark matter only simulations also reveal the existence of massive structures within the cluster-scale halos, that could also be interpreted as free-floating black holes \cite{Banik2019}. 
\review{Two mechanisms could originate the wandering SMBH population. First, infalling galaxies is the main mechanism of clusters growth and only a fraction of the galaxy survive. While being completely stripped out of their stars in their infall from tidal friction \citep{Wu2013,Haggar2022}, the remaining central black hole can survive and stay within the halos without enough stars around it to be directly detected. A second mechanism is the the ejection of a SMBH due to recoil as described in \citet{Paynter2022}.
}
To implement our solution we radially distribute wanderers following a log-normal distribution as a function of projected cluster-centric radius as reported in \cite{Ricarte2021a} (see Figure 6 in their analysis) with a mean at 0.1 R$_{200}$ (radius at 200 times the critical density of the Universe) of the host halos. In this work, we use a cluster host halo of $5.10^{14}\Msunmath $ and a R$_{200}=1.5\,$Mpc. In the high-resolution simulated cluster Romulus-C studied by \cite{Ricarte2021a}, they report finding over 1600 wanderers distributed within the virial radius. In accordance with this demographic, we randomly draw 1000 locations for our wanderers.

We repeat this mock cluster simulation a 1000 times following the mass and positions described above. 
To mimic realistic scenarios, we choose to limit ourselves to a mass range of $10^6 \Msunmath$ to $10^{10} \Msunmath$ for the SMBH masses. Following these constraints, we numerically computed the probability for the presence of massive SMBH. There is 3.7 \% probability that at least one SMBH wanderer with a mass above $10^8$ \Msun\, acts as a lens and this probability drops to 0.07\% to have at least two SMBHs above $10^8$ \Msun. The probability for at least 1 SMBH to be more massive than $10^9$ \Msun\, is about 0.02\% (or 1 per 5000 clusters).

\section{SMBH lensing signatures in specific multiple image configurations}
\label{sec:obs-effect}

Observationally detectable lensing signatures of a SMBH or for a compact dark matter halo without a sufficiently luminous counterpart exhibit a large variety of image configurations. The most obvious cases emerge when a SMBH is near a critical curve or when its impact is boosted by significant amplification from the larger-scale cluster host halo. In the following section we summarise the five distinct categories of potential observational signatures of lensing by a SMBH:
\begin{enumerate}
    \item Image splitting due to alignment of the SMBH inside the cluster critical curve.
    \item Change in the observed flux of one of the lensed images. 
    \item Apparent skewing in the light profile (or velocity field) and an apparent break in symmetry.
    \item Apparent increase in the size of a single lensed image
    \item Apparent disappearance of an internal image
\end{enumerate}
We detail each of these categories further in the following subsections.

\subsection{Image splitting due to alignment of the SMBH inside the cluster critical curve.}

\begin{figure*}
    \centering

    \includegraphics[width=0.35\linewidth]{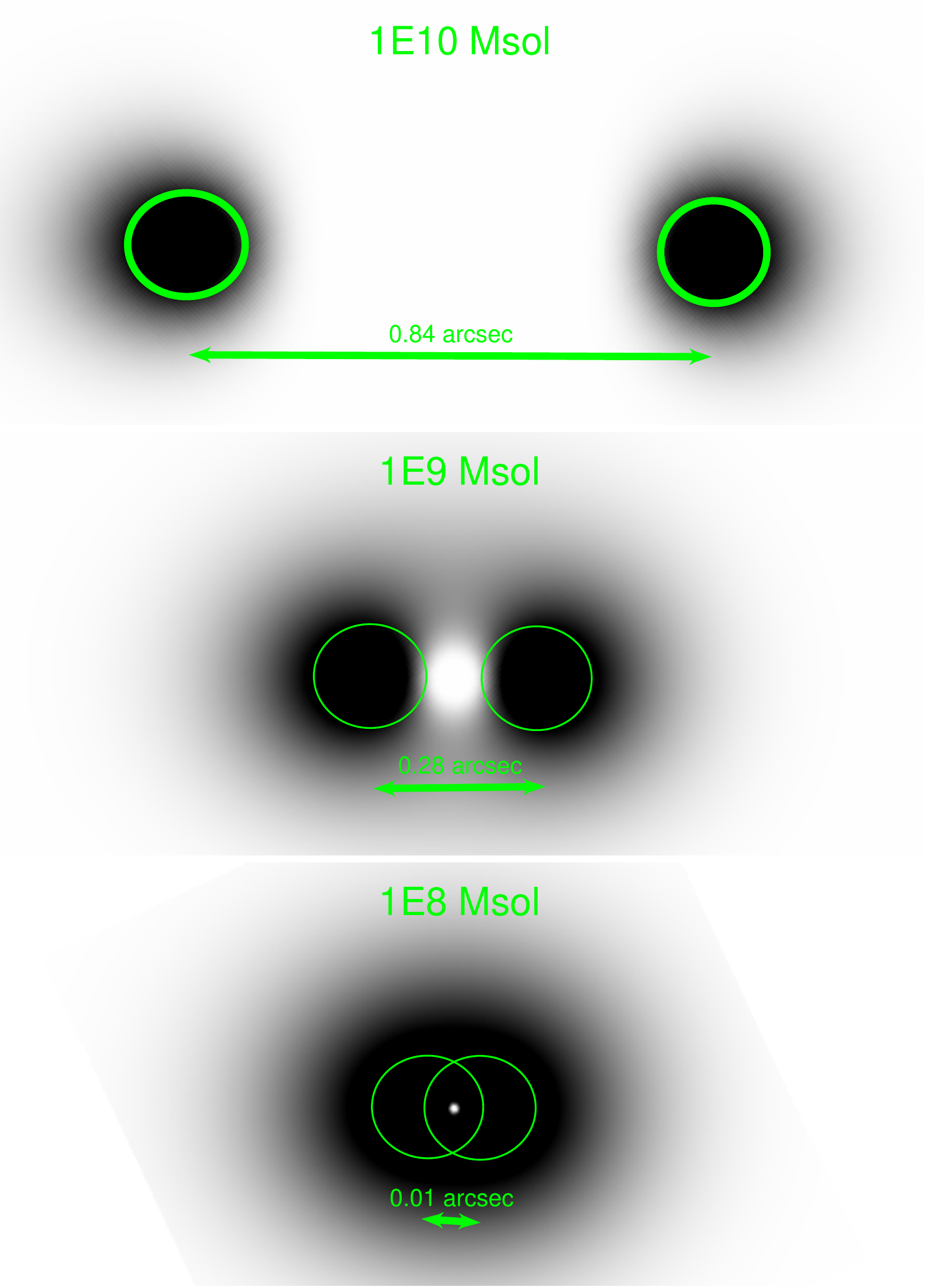}
    \includegraphics[width=0.6\linewidth]{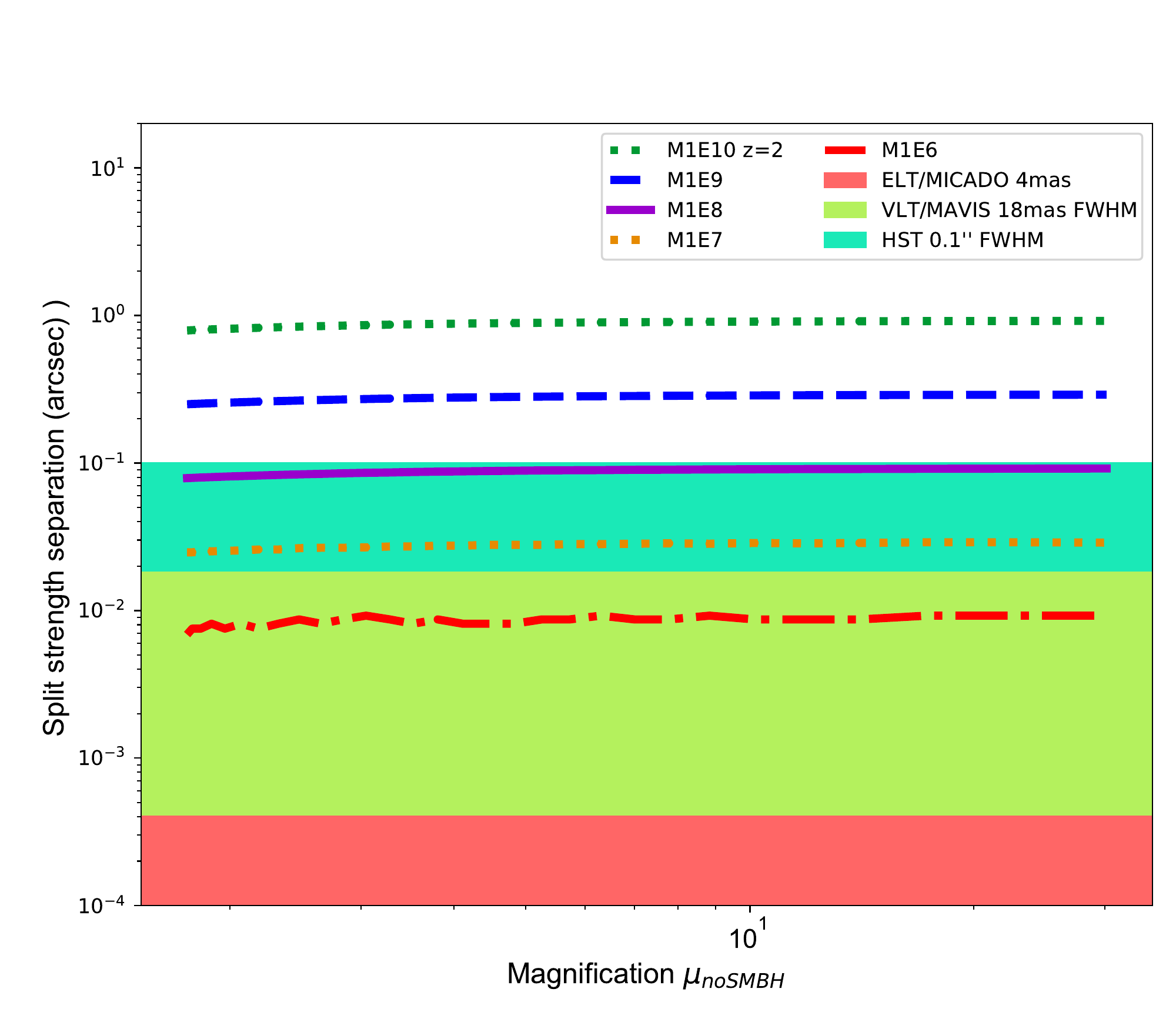}
    \caption{This figure shows the splitting power of an SMBH. \textit{Left:} Illustration of the splitting of the image 
    \review{by a SMBH in a z=$0.2$ cluster of a $\sigma_G=$0.1\arcsec gaussian source at z$=$2 (corresponding to 0.85\,kpc)}, from top to bottom the mass of the SMBH is varied ranging from $10^{10}\Msunmath$, to $10^{9}\Msunmath$, to $10^{8}\Msunmath$. \textit{Right:} separation of the image split plotted  as a function of the underlying cluster magnification for different masses of the SMBH. We see that while a massive SMBH can be detected easily with HST resolution data, even the effect of a SMBH that is a few time $10^7$ \Msun\, is potentially detectable providing there is perfect alignment. We also highlight here that the future VLT/MAVIS instrument will reach 18 mas resolution and the promised sampling of ELT/MICADO can reach 4 mas, bringing the resolution down to levels that might permit capabilities to detect SMBH lensing in cluster environments.}
    \label{fig:separation}

\end{figure*}

The presence of an SMBH introduces a perturbation in the lensing configurations as noted. If the image of the lensed source is exactly aligned with the SMBH the resulting image is split. The perfect alignment represents the maximum splitting that SMBH can generate, that we refer to that later as the splitting power. Figure \ref{fig:separation} illustrates the splitting power of SMBHs as a function of their mass. This splitting power is mainly driven by the actual mass of the 
SMBH, other contributing factors are redshift, a smaller variation appears with increasing redshift of the source (up to 25\% for redshift 10); the underlying magnification (up to 10\% for an order of magnitude change in magnification) and the cluster induced shear(up to 25\%). From Figure\,\ref{fig:separation}, we clearly see that an SMBH of more than $10^{8}\,\Msunmath$ has a large enough splitting power to be detected by \HST. However, we note here that the size of the caustic curve in the source plane is small. For example, at $z = 2$, a source in alignment with a $10^{8}\,\Msunmath$ SMBH, needs to have distinct feature emitting light smaller than $\sim$100\,pc. If the light source is homogeneous and covers a larger area, an observer cannot identify the two separate components. This would result in observations similar to the one detailed in \ref{sec:elong-asym}.

\subsection{Change in the observed flux of one of the lensed images}

\begin{figure*}
    \centering
    \includegraphics[width=\linewidth]{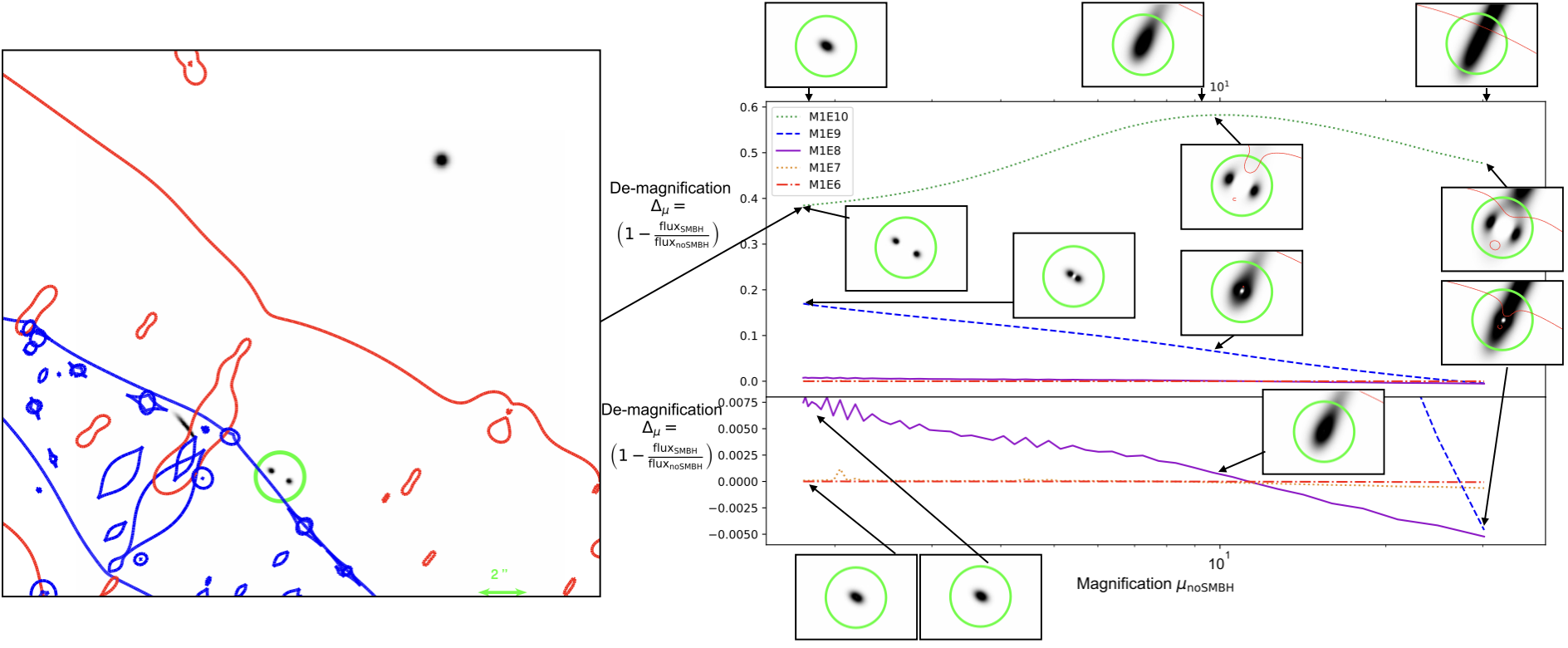}
    \caption{\review{Simulation of a $z=2.0$ $\sigma_G=$0.1\arcsec gaussian source at z$=$2 (corresponding to 0.85\,kpc) lensed by a SMBH in a z=$0.2$ cluster.}
    \textit{Left:} Simulation of a source at $z=2.0$ that appears to be de-magnified by a $10^{10}$ \Msun SMBH, with its location marked as the green circle. The caustic and critical curves are shown in blue and red respectively. The underlying magnification provided by the cluster is assumed to be a factor of 1.75 \textit{Right:} The de-magnification induced by the presence of an SMBH within a circle of a 1 arcsecond radius as a function of the underlying cluster-induced magnification. The de-magnification factor measures how much flux is lost compared to the case without the lensing effect of the SMBH. No instrumental effects have been added for these plots. Further explorations of the de-magnification as a function of the underlying cluster magnification are available at: \url{https://sites.google.com/view/guillaume-mahler-astronomer/paper-animation/paper-animation-demag}}
    \label{fig:demag}
\end{figure*}

In the case of perfect alignment between the SMBH and the lensed sources, the produced image can be split. Using the same mock simulation, we measure the flux within a 1.0 arcsecond radius aperture around the image of the source. We find that the apparent change in flux depends greatly on the mass of the SMBH and the underlying magnification of the host cluster. We attempt to quantify the  de-magnifying effect and derive the decrements in the flux ratio. Figure \ref{fig:demag} shows the de-magnification as a function of magnification boost from the cluster ranging from 1.7 to 30, when the SMBH and image remains well within the cluster's internal critical curve. 

We note that the underlying magnification is more important for a massive SMBH for low magnifications. We interpret this as a consequence of the regime where the underlying properties of the SMBH dominates. For underlying magnifications $\gtrsim$ 9 the detected flux variation converges to only 1\% for all masses and magnifications.
We note in our simulation that for masses below $10^8$ \Msun, the effect on the flux is the nearly indistinguishable from the case with no SMBH. Only at much higher resolutions, in a situation where the source is smaller than the caustic curve of the SMBH, does the effect of such a configuration become noticeable. We don't anticipate current and near future facilities to be able to have such resolving capabilities but the main challenge to measure this effect might well be the accuracy of our mass modelling techniques.

\subsection{Induced asymmetry with the skewing of the light profile and velocity field}

\begin{figure}
    \centering
    \includegraphics[width=\linewidth,trim={23cm 0 23cm 0 },clip]{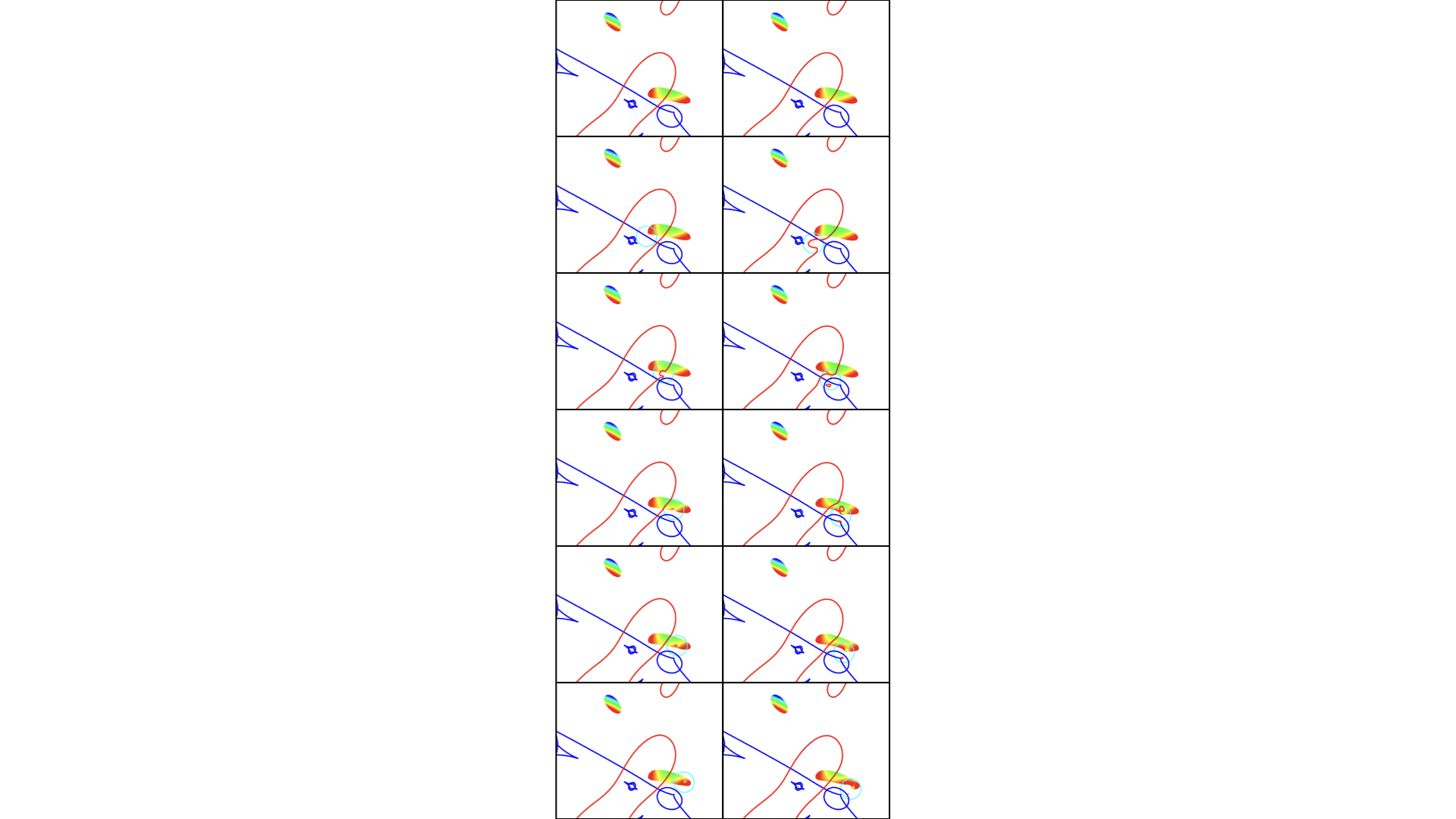}
    \caption{Light distribution distortion of a lensed sources \review{modelled as a disk of 60 mas radius at z$=$2 (corresponding to 0.5\,kpc)} as a function of the SMBH location and mass \review{in a cluster at z$=$0.2}. The 0.3\arcsec radius (cyan circle) marks the location the SMBH. The blue and red curves mark the caustics and critical curves respectively. The x-axis locations are the same for each panel and the y-axis locations for the left columns (10$^{8} \Msunmath$) are higher than for the \review{right columns (10$^{9} \Msunmath$) }for the effect of the  SMBH to be seen. We notice how the presence of the SMBH distorts the image gradient. This would be visible either in the velocity field of the galaxy or potentially in looking at an unexpected  symmetry between two images of the same sources, more easily identifiable in clumpy galaxies. Further exploration of the distortion as a function of SMBH location for a range of SMBH masses are available at \url{https://sites.google.com/view/guillaume-mahler-astronomer/paper-animation/paper-animation-warp}
    }
    \label{fig:warp}
\end{figure}

The presence of SMBH wanderers aligned with sources will influence observed lensing configurations, detectable in the case of a source with enough resolution elements or a sharp gradient in color, such as would be seen in a composite image of the velocity field. The presence and proximity of an SMBH can be detected by looking at the light distribution as shown in Figure\,\ref{fig:warp}. Less massive SMBHs have a smaller area of influence and therefore need to be closer to the image to produce detectable effects. The detection of the presence of a SMBH is optimal for the configuration showed in Figure\,\ref{fig:warp} where the lensed image has a merging pair configuration and the SMBH is close to the critical curve. This warping of the image is also very similar to the effect induced by sub-halos seen in galaxy-galaxy lensing events (e.g. \citealt{Hezaveh2016}). This effect might be the most common effect the SMBHs produce viz-a-viz lensing. At a constant underlying magnification, the strength of the warping is mainly affected by the distance between the lensed image and the SMBH. For instance, a $10^{10}\Msunmath$ SMBH would have a significant effect  up to $\sim$0.5\arcsec away from the arc, a $10^{9}\Msunmath$ SMBH up to $\sim$0.3\arcsec, and a $10^{8}\Msunmath$ SMBH up to $\sim$0.1\arcsec.

\subsection{Apparent increase in the size of one of the lensed images of a multiple}
\label{sec:elong-asym}

\begin{figure}
    \centering
    \includegraphics[width=\linewidth]{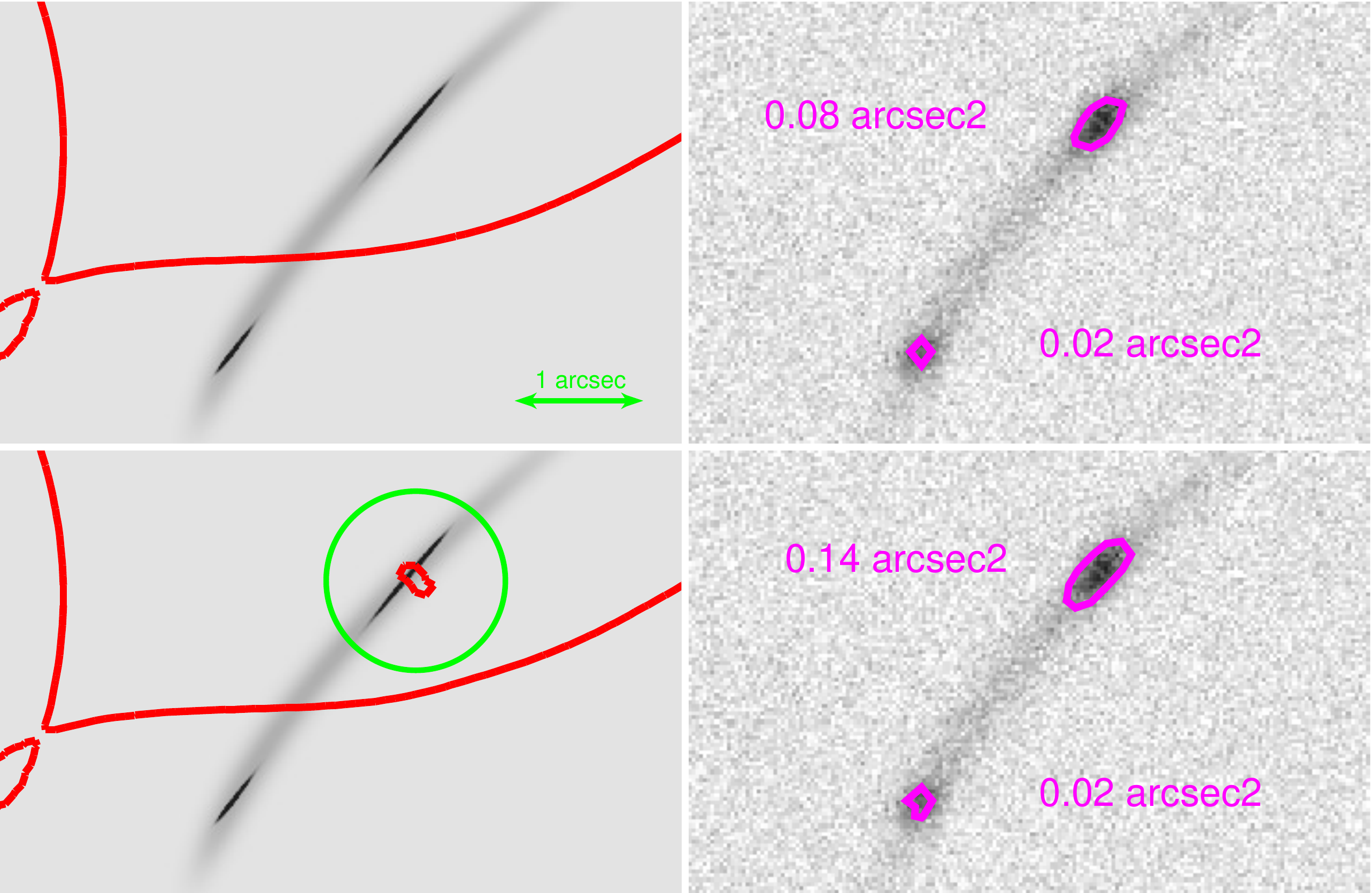}
    \caption{Simulation involving a two-component source: a compact \review{($\sigma_G=$1mas)} and a larger \review{($\sigma_G=$7mas)} gaussian light distribution \review{at z$=$2 lensed by a cluster at z$=$0.2}. 
    Top panels show the configuration without a SMBH and bottom panels with a $10^{8} \Msunmath$ SMBH at the location of the 1\arcsec radius (green circle). Left panels show the simulation at infinite resolution and the right panels show observations similar to what ACS/HST would observe (0.1\arcsec FWHM PSF, 30 mas pixel scale, with noise scaled as a 10\% of the maximum flux value). We see instrumental effects result in a larger area for the compact component, rendering it detectable.
     } 
    \label{fig:elong-asym}
\end{figure}

As discussed before, the more massive a SMBH the stronger impact it generates on the overall lensing configuration. In addition, for real observational facilities, taking into account additional effects such as the impact of the point-spread function, and pixel size or noise level also serve decrease our ability to precisely identify SMBH induced lensing configurations. However, even after instrumental effects are taken into account, the light distribution might reveal information to hint the presence of a SMBH. The primary effect would be to increase the apparent size of the image and this too preferentially in the lensing configuration where the SMBH influences an image outside the main cluster critical curve. Figure \ref{fig:elong-asym} shows an example of the larger flux distribution induced by an SMBH (M$_{SMBH}=10^{8} \Msunmath$). The source is modeled here as a double Gaussian to simulate the compact and broad emission components. The SMBH is placed outside the critical curve in this case. We note that in this example the area of the principal compact emission knot almost doubles. With a good prior understanding of the source light distribution such a measurement can lower the detectability threshold of SMBHs down to M$_{SMBH}=10^{8} \Msunmath$ with \HST specifications.

\subsection{Apparent disappearance of the internal image}
\label{sec:disa}

\begin{figure}
    \centering
    \includegraphics[width=\linewidth]{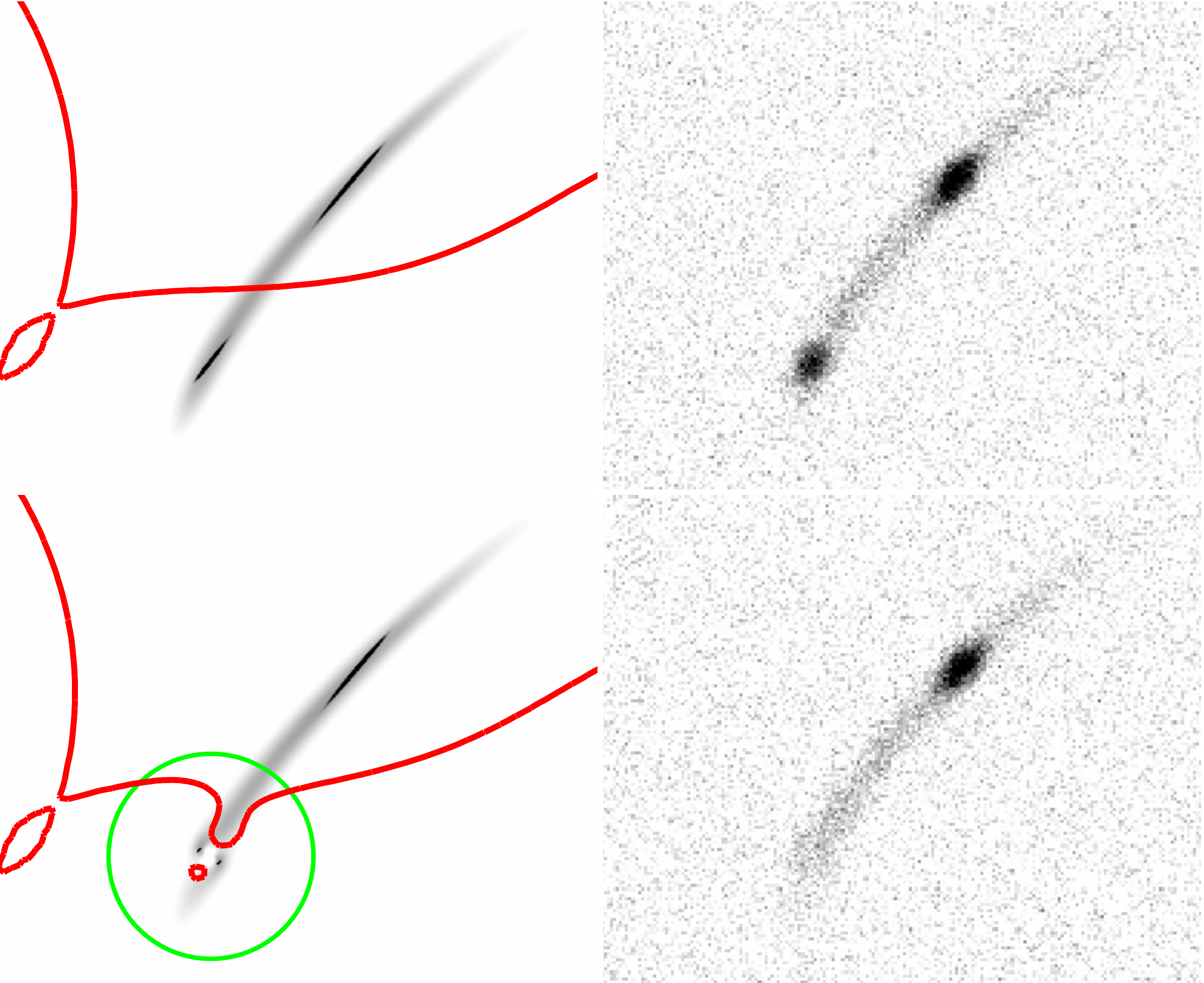}
    \caption{Simulation involving a two-component source: a compact \review{($\sigma_G=$1mas)} and a larger \review{($\sigma_G=$7mas)} gaussian light distribution \review{at z$=$2 lensed by a cluster at z$=$0.2}. 
    Top panels show the configuration without a SMBH in the cluster and bottom panels with a $10^{9} \Msunmath$ SMBH at the location of the 1\arcsec radius (green circle). Left panels show the simulation at infinite resolution and the right panels show the observation mimicking ACS HST observations (0.1 \arcsec FWHM PSF, 30 mas pixel scale, with noise scaled as a 10\% of the maximum flux value). We see the instrumental effects result in the apparent disappearance of one of the images. The mass of the SMBH and the size of the source both contribute to this effect. We note that it prominently appear when the SMBH is near perfect alignment on of the image inside the cluster critical curve. Instrumental effects can more easily smooth the light distribution inside the cluster critical curve because the SMBH only split the image in two (instead of four outside)} 
    \label{fig:rm-in}
\end{figure}

As discussed previously, the splitting power of an SMBH can separate images and therefore redistribute the light. When an SMBH is aligned almost exactly with the image of a lensed system inside the cluster critical curve the resulting effect is to separate the image into two ( or into four if it happens to be outside the cluster critical curve). If the image is only split into two, instrumental and observational effects can affect their detectability resulting in the actual disappearance of one of the images. Figure \ref{fig:rm-in} highlights such situation. The PSF convolution and noise induced by the instruments, smooth the signal and make it harder to detect, giving an impression of the disappearance of the image and hence a break in the ''typical'' image symmetry.

The apparent disappearance of the inner counter-images is the easiest case to detect in the upcoming large image survey data but this requires as noted a very tight alignment between the image and the SMBH and is therefore, expected to rare. In addition, it is degenerate with the large scale cluster model and detailed source morphology.

\section{Observational tests}\label{sec:realcase}

In this section, we explore for evidence of an SMBH in currently available lensing observations. While we did not find any clear-cut evidence of SMBH lensing in our current fairly exhaustive search of peculiar lensing configurations, we found two cases where a wandering SMBH might provide a potentially plausible explanation to account for what is seen. As mentioned clearly at the start of our investigation, it is extremely challenging to pinpoint the role of an SMBH in any observed configuration as it is degenerate with the effect produced by a dark sub-halo. 
We want to stress here that tweaking other models not including SMBHs are also able to account for the observed data. However, the parameter space available to interpret the data is still open and it is in this context that we explore the possible role for SMBHs as lenses.

\subsection{Asymmetrical lensing configuration in SGAS\,J003341.5$+$024217}

\begin{figure*}
    \centering
    \includegraphics[width=\linewidth]{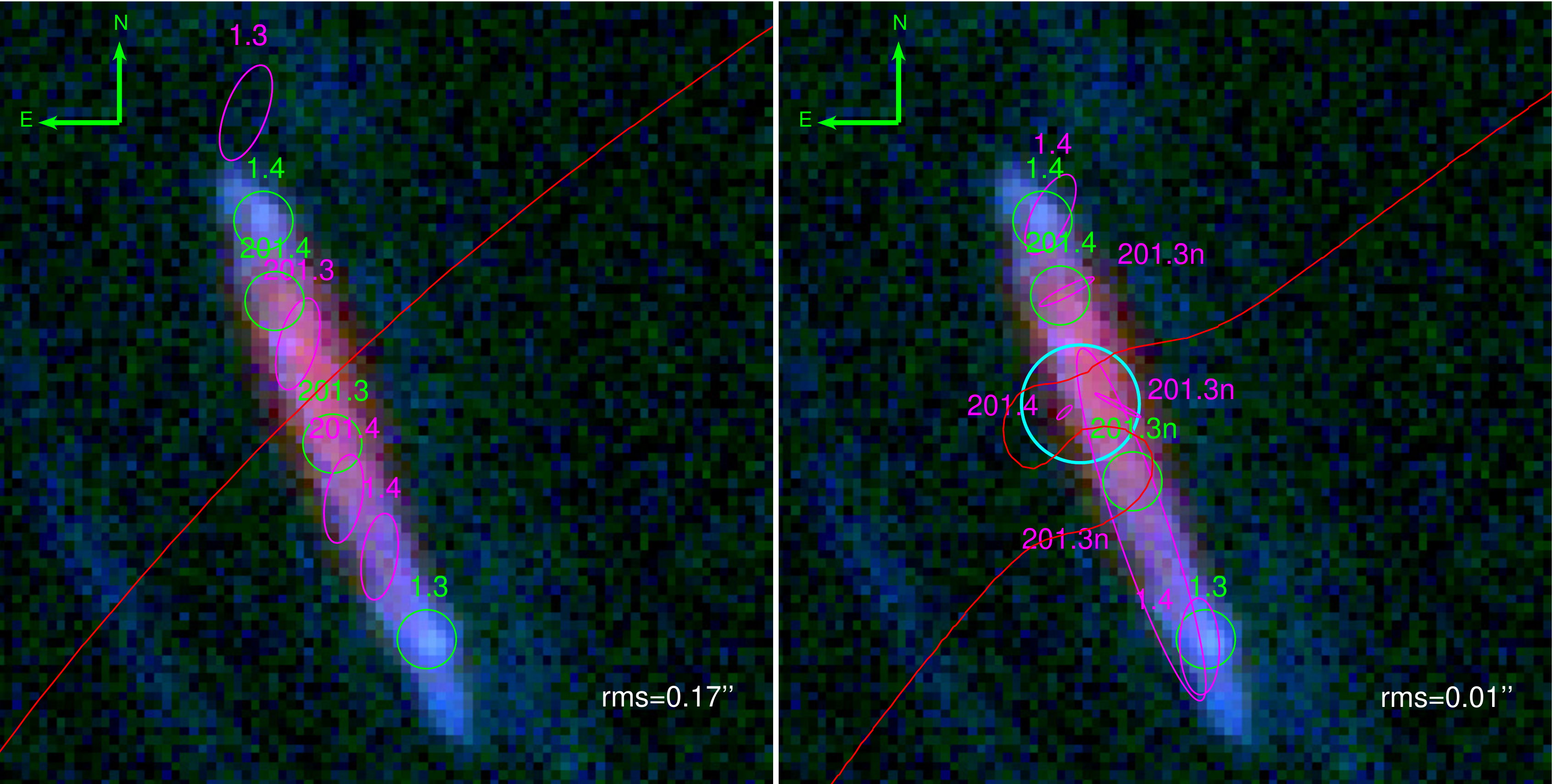}
    \caption{Previous analysis report that SGAS\.J003341.5$+$024217 show peculiar unexplained asymmetry \citep{Fischer2019}. \textit{Left:} Original model; The lensing configurations is a classically describe as a merging pair, the critical curve (red curve) is passing through the centrally located red part of the system but the two extreme emission knots are not symmetrically located on each side of the critical curve as expected. \textit{Left:} Alternative modelling including a SMBH at the location of the critical curve alleviate this tension. Our models that include a $4.7~10^{8}$\Msun SMBH at the location of the cyan circle (0.2 \arcsec radius) perturbed the lensing critical curve to reestablish symmetry on both side of the curvy line. The green circle represents the emission knot on the image and the magenta circle are the corresponding predictions from the lens model. } 
    \label{fig:sdss0033}
\end{figure*}

The observational configuration of the arc shown in Figure\,\ref{fig:sdss0033} is akin to the case referred in sub-section \ref{sec:elong-asym}, though SMBH is required to be located on top of the critical curve. \cite{Fischer2019} report the asymmetrical lensing configuration for an arc at $z=2.39$ merging on the critical curve. To model this case, here we include a wandering SMBH as a perturbation to account for the asymmetrical shape of this arc as shown in Figure\,\ref{fig:sdss0033}. The addition of the lensing perturbation from a SMBH allows us to reduce the rms positional error in the current best-fit mass model of the system from 0.17\arcsec that does not include this additional degree of freedom of an SMBH, down to $ 0.01$\arcsec including the SMBH. This improvement including this additional degree of freedom is real, as it is reflected in the value of the reduced $\chi^2$ as well. Therefore, the inclusion of a wandering SMBH offers a plausible explanation for constructing a robust model for the observation.

\subsection{Disappearance of lensing symmetry in RX\,J1347.5-1145}

\begin{figure}
    \centering
    \includegraphics[width=0.9\linewidth]{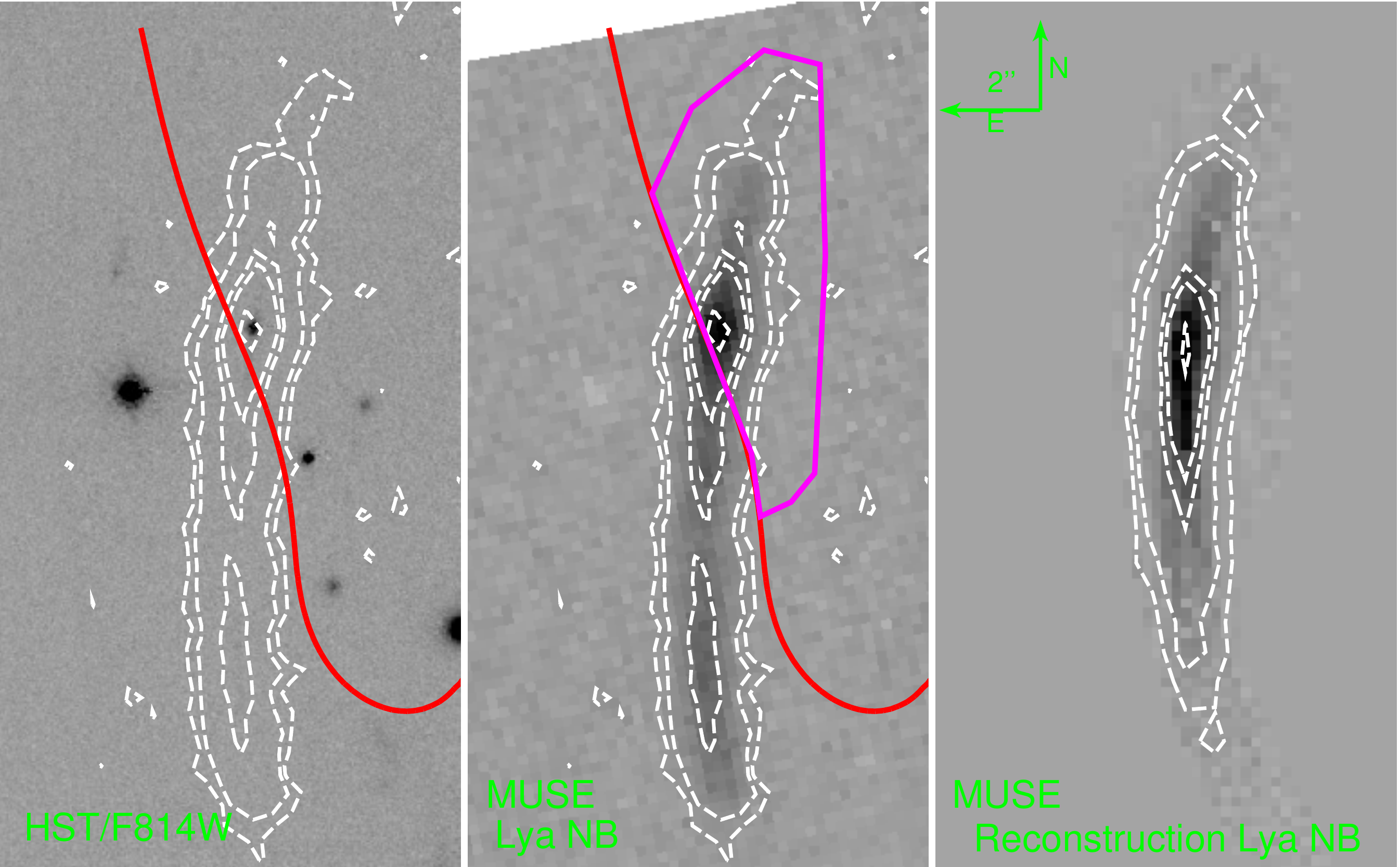} 
    \includegraphics[width=0.9\linewidth]{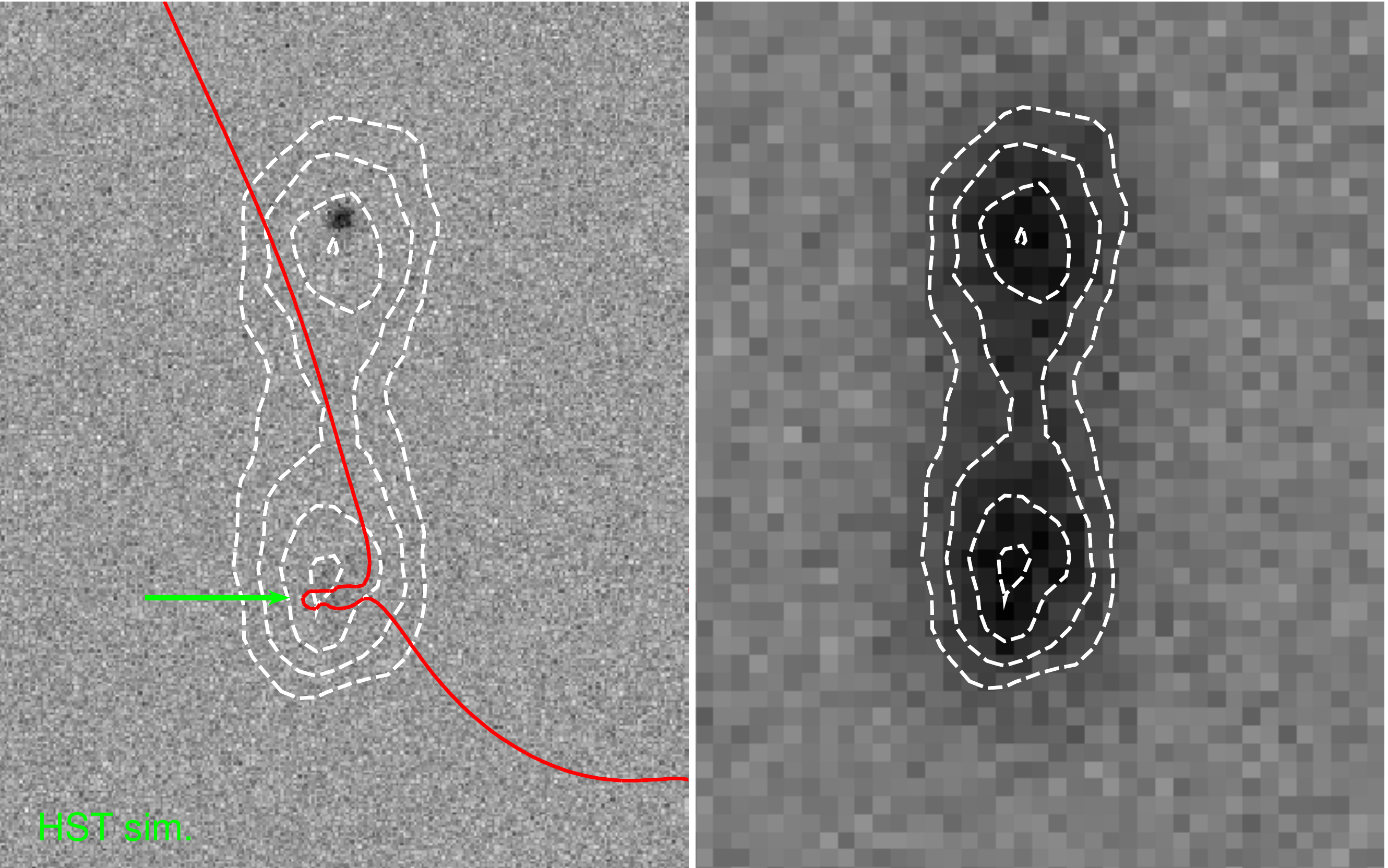} 
    \caption{Alternative explanation for the peculiar lensing configuration in the multiply imaged system in RX\,J1347.5-1145. \textit{Top}: The panels show system 25 (as reported in \citealt{Richard2021}) in the HST/F814W band (left), MUSE narrow band images centered on the Lyman alpha sources (middle) and a lens re-projection of the top part of the MUSE NB image Lyman alpha emission. This re-projection consist of selecting the pixels in the magenta polygon on the top left panel, sending them to the source plan and projecting them back to the image plan, showing the flux distribution assume with this lens model. The bottom part of the system does not match the expected lensing symmetry highlighted by the critical curve at the redshift of the images shown in red. \textit{Bottom}: Our model offer alternative explanation of the lensing configuration: we slight modify the large scale mass component to shift the critical curve to lie in between the two Lyman-alpha peak. In addition we include a $5\,10^{8}$\Msun SMBH at the location of the green arrow showing as a "kink" in the critical curve. \review{We model the continuum emission from HST with a simple gaussian with $\sigma_G=1mas$ and the Lyman-alpha emission  with a elliptical gaussian with $\sigma_G=18mas$ and $\sigma_G=18mas$ for the semi-major and semi-minor axis at a 90 degree angle from the horizontal direction. This peculiar configuration accounts for the disappearance of the continuum peaky emission due to the "splitting" of the image as noted in \ref{sec:obs-effect}. The diffuse emission, because of the combined effect of it's larger size and larger seeing mocked here with a FWHM = 0.5\arcsec), appear slightly more elongated on the lower portion of the diffuse emission, as in the observation.}
    }
    \label{fig:rxj1347}
\end{figure}

\citet{Richard2021} report a peculiar lensing configuration in a massive cluster observed with the combination of HST and MUSE. They identify unique compact continuum emission identified in HST and extended gas emission detected in MUSE that shows a large tail (see their Figure 12 or top panels of Figure \ref{fig:rxj1347}). With the lensing configuration from the authors' published model, we would expect similar size for the extended tail on both sides of the critical curve. Here, we propose a different interpretation of the observations resembling the configuration studied in section \ref{sec:disa} where there is an apparent missing counter image. As reported earlier, a missing counter-image results when the source falls inside the cluster critical curve. Starting with the published models from \citealt{Richard2021}, we manually tweaked the main cluster model parameters to move the cluster critical curve to the middle of the lobe of the extended emission accomplished using an SMBH. We simulated the emission with two Gaussian distributions on top of each other, to represent the compact continuum emission
\review{ with a circular gaussian with $\sigma=1mas$ and the Lyman-alpha emission with a elliptical gaussian with $\sigma_G=18mas$ and $\sigma_G=18mas$ for the semi-major and semi-minor axis at a 90 degree angle from the horizontal direction. For a source at z$=$4.0840 \citep{Richard2021} this correspond to a 7\,pc and 126\,pc for the continuum and Lyman-alpha emission respectively.} 
to mimic the Lyman alpha emission of the observed arc. The northern part of the extended emission is inside the cluster member critical curve and shows the compact continuum emission. We placed an SMBH on the southern part of the arc close enough such that cluster critical curves merge, effectively placing the SMBH inside the main cluster critical curve. This leads to the splitting of the compact emission component on the southern part of the arc. The inclusion of observational effects causes the extended light emission to be smeared into a more elongated shape, reproducing the observation. We, therefore, conclude  that with the addition of a $5.10^8\Msunmath$ SMBH and a simple model for the source light distribution we can qualitatively reproduce the observations. The implication of lensing by a SMBH offers an alternative, viable interpretation. While this is by no means a claim for the detection of the lensing effect of a SMBH, it is simply a demonstration that inclusion of such a perturber offers a neat and natural explanation for the observed asymmetry seen in RX\,J1347.5-1145.

\section{Conclusions}\label{sec:ccl}

Our analysis explores the observational lensing signatures of SMBHs in galaxy clusters. We model the mass of SMBHs as compact dPIE profiles and study their effect both as centrals hosted by cluster member galaxies as well as free-floating wanderers in the larger cluster environment. The presence of a population of wandering SMBHs in clusters has been recently claimed by \citealt{Ricarte2021a} from the analysis of high-resolution cosmological simulations. Although the lensing effect of a SMBH is small, it imprints a discernible signature in unusual  lensing image configurations, as summarised below:   
\begin{itemize}
    \item The presence of a central SMBH primarily affects the radial critical curve of its host, resulting either in its disappearance or by its splitting. Hint of a split critical curve would be a clear evidence for the presence of a central SMBH, but its detection remains challenging due to the overlap of the radial critical curve and the light of the lens itself. Large high-resolution radio surveys might be able to support a concerted search in the future due to the faintness of the emission from the lens at these wavelengths. 
    \item The shear induced by central SMBHs is statistically impossible to disentangle from the effect of a change in the density profile at the core of their host galaxies.
    \item Wandering SMBHs can cause the disappearance of counter-images. This effect would be the most obvious signature, however, it is also degenerate with the lensing properties of the larger scale cluster mass model as well as complexities in the intrinsic morphology of background sources.
    \item Outside the cluster critical curve, the typical effect of a wandering compact SMBH would result in the asymmetric elongation of one of the images. This might be much a more common occurrence but to conclusively confirm the presence of the SMBH, a detailed lens model and careful source reconstruction is required to show this anomaly in the reconstructed image. 
    \item The changes induced by wandering SMBHs in the light distribution of lensed sources can be tracked in the case of a obvious change in the symmetry of the field, ideally traced where clear gradients exist, such as in the velocity field of a lensed galaxy. Observing this feature requires robust mapping of the true velocity field of the source.
\end{itemize}

We attempt in this analysis to reconcile previously reported peculiar lensing configurations in SGAS\.J003341.5$+$024217 and RX\,J1347.5-1145 by including  perturbations induced by an SMBH into the mass models. With these new composite mass models, we do not claim detection or discovery of the presence of a SMBH, but rather demonstrate that a class of models with SMBHs (M$_{\rm SMBH}=4.7\,10^8\Msunmath$) might be able to alleviate some of the previously noted tensions in modeling these unusual lensing configurations. 

We report that the ideal configurations that would reveal the presence of an SMBH without doubt would be the appearance of a two-image split inside the cluster critical curves. We have not been able so far to identify such a split in observational data currently in hand. In the future, two complementary strategies can be pursued to discover wandering SMBHs. Future large surveys by the Rubin LSST Observatory, \textit{Euclid}, or \textit{the Nancy Grace Roman Telescope} will offer orders of magnitude more targets and will likely therefore uncover rare events of massive SMBH alignment. The second strategy would be to point extremely high resolution cameras on giant arcs and/or multiply-imaged systems of clumpy galaxies to push down on the granularity of the mass distributions and hence the mass threshold of detectable SMBHs.

\section*{Acknowledgements}

GM acknowledges funding from the European Union’s Horizon 2020 research and innovation programme under the Marie Skłodowska-Curie grant agreement No MARACHAS - DLV-896778. PN acknowledges the Black Hole Initiative (BHI) at Harvard University, which is supported by grants from the Gordon and Betty Moore Foundation and the John Templeton Foundation. PN acknowledges useful conversations with Angelo Ricarte and Michael Tremmel. MJ is supported by the United Kingdom Research and Innovation (UKRI) Future Leaders Fellowship `Using Cosmic Beasts to uncover the Nature of Dark Matter' (grant number MR/S017216/1).

\section*{Data Availability}

The data underlying this article are available in the article and in its online supplementary material. \url{https://sites.google.com/view/guillaume-mahler-astronomer/paper-animation}



\bibliographystyle{mnras}
\bibliography{example} 




\appendix
\section{Ellipticity of the cluster halo}
\label{sec:app-ell}
\review{
Following our analysis on lensing configurations for different profil and masses of SMBH as described in \ref{sec:BCGcentral} We quickly document here two cases of the influence of the cluster scale halo ellipticity on the lensing configurations for a combinations of three mass. A cluster halo simulated as an NFW, parameterised as $c=7$, r$_s=200$\,kpc giving a total mass of M$_{200}=3.8~10^{14}\Msunmath$), a host galaxy parameterised with a dPIE (r$_{{\rm core}}=0.1$\,kpc, r$_{{\rm cut}}=50$\,kpc, and SMBH with two masses M$_{\rm SMBH}= 1~10^{8}$\Msun and M$_{\rm SMBH}= 1~10^{10}$\Msun. We can notice that the critical at the center is able to split at an earlier mass than the cases report in \ref{fig:BCG-panel}, if the cluster scale halo show a high enough ellipticity.
}
\begin{figure}
    \centering
    \includegraphics[width=\linewidth,trim={22cm 0 21cm 0},clip]{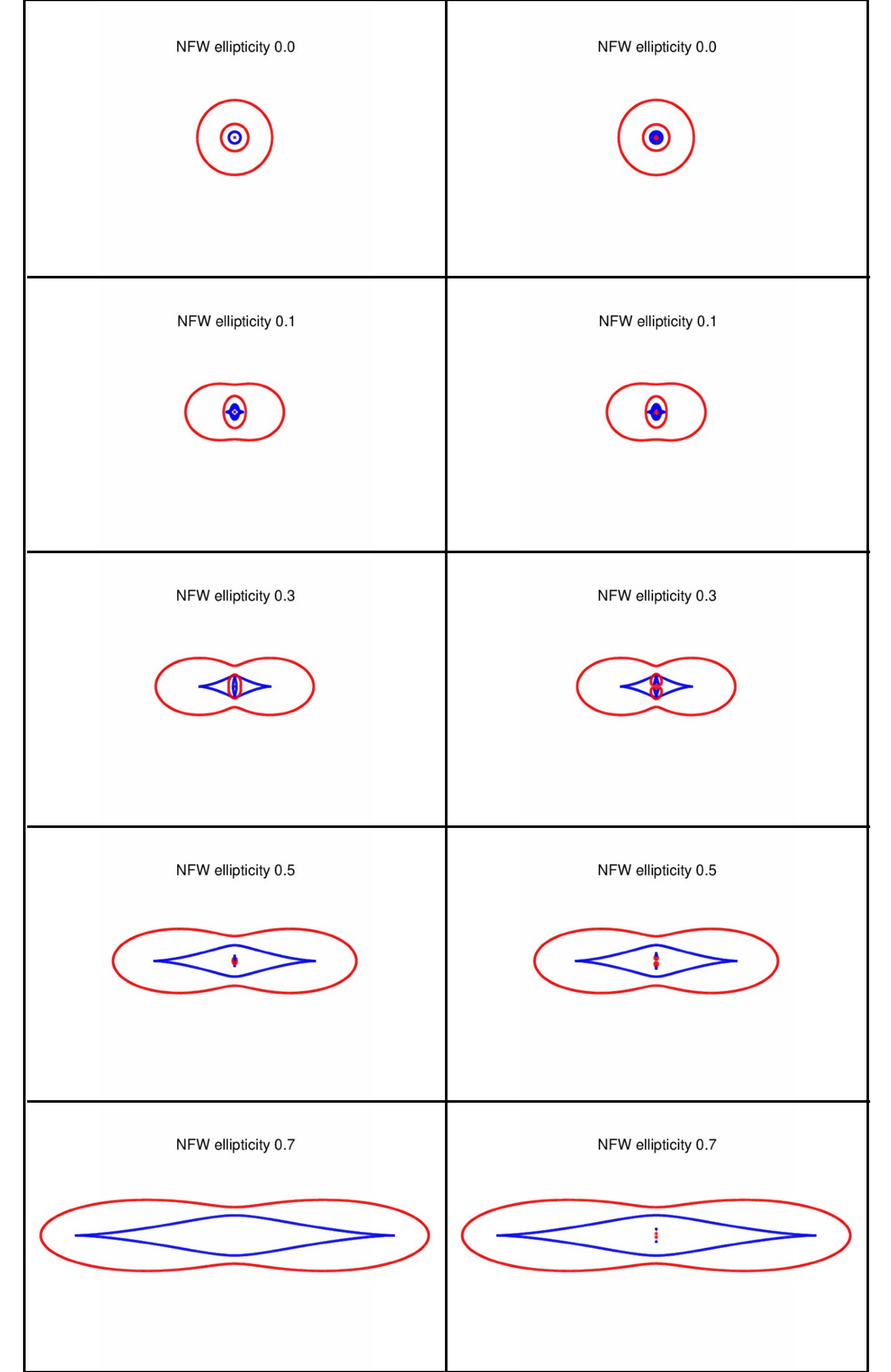}
    \caption{
    \review{Lensing configurations for a $10^{12}\Msunmath$ galaxy at the center of an NFW profile with varying ellipticity for two different mass of central SMBH. A SMBH mass of M$_{\rm SMBH}= 1~10^{8}$\Msun for the left column and M$_{\rm SMBH}= 1~10^{10}$\Msun on the right column. The NFW profile characterized by the following parameters: $c=7$, r$_s=200$\,kpc and a total mass of M$_{200}=3.8~10^{14}\Msunmath$. The host galaxy is parameterise with a dPIE (r$_{{\rm core}}=0.1$\,kpc, r$_{{\rm cut}}=50$\,kpc, $\sigma_{\rm{dPIE}}=135$km\,s$^{-1}$) and has a total mass $10^{12}\Msunmath$.
    We can see that the ellipticity plays a role in the lensing configuration and can lower the mass when the internal critical curve split into two part as detailed in \ref{sec:BCGcentral}. 
    Each box has a length of 85\arcsec aside, corresponding to 280\,kpc at z $=0.2$. 
    }
    }
    \label{fig:ellNFW}
\end{figure}


\bsp	
\label{lastpage}
\end{document}